\documentclass[sigconf]{acmart}

\AtBeginDocument{%
  }

\setcopyright{acmlicensed}
\copyrightyear{2025}
\acmYear{2025}
\acmDOI{XXXXXXX.XXXXXXX}
\acmConference[MM '25]{ACM International Conference on Multimedia 2025}{October 20--24, 2025}{Dublin, Ireland}
\acmISBN{978-1-4503-XXXX-X/2018/06}

\usepackage{multirow}
\usepackage{subcaption}
\usepackage{array}
\usepackage{algorithm,algpseudocode}
\newcolumntype{R}[1]{>{\raggedleft\arraybackslash}p{#1}}

\begin{document}
\title{PCDiff: Proactive Control for Ownership Protection in Diffusion Models with Watermark Compatibility}

\author{Keke Gai}
\affiliation{%
 \institution{Beijing Institute of Technology}
 \country{China}}
\email{gaikeke@bit.eud.cn}

\author{Ziyue Shen}
\affiliation{%
 \institution{Beijing Institute of Technology}
 \country{China}}
\email{3220242027@bit.eud.cn}

\author{Jing Yu}
\affiliation{%
 \institution{School of Information Engineering,
Minzu University of China.}
 \country{China}}
\email{jing.yu@muc.edu.cn}

\author{Liehuang Zhu}
\affiliation{%
 \institution{Beijing Institute of Technology}
 \country{China}}
\email{liehuangz@bit.eud.cn}

\author{Qi Wu}
\affiliation{%
 \institution{University of Adelaide, Adelaide}
 \country{Australia}}
\email{qi.wu01@adelaide.eud.au}

\renewcommand{\shortauthors}{Trovato et al.}

\begin{abstract}
With the growing demand for protecting the intellectual property (IP) of text-to-image diffusion models, we propose PCDiff---a proactive access control framework that redefines model authorization by regulating generation quality. At its core, PCDIFF integrates a trainable fuser module and hierarchical authentication layers into the decoder architecture, ensuring that only users with valid encrypted credentials can generate high-fidelity images. In the absence of valid keys, the system deliberately degrades output quality, effectively preventing unauthorized exploitation.Importantly, while the primary mechanism enforces active access control through architectural intervention, its decoupled design retains compatibility with existing watermarking techniques. This satisfies the need of model owners to actively control model ownership while preserving the traceability capabilities provided by traditional watermarking approaches.Extensive experimental evaluations confirm a strong dependency between credential verification and image quality across various attack scenarios. Moreover, when combined with typical post-processing operations, PCDIFF demonstrates powerful performance alongside conventional watermarking methods. This work shifts the paradigm from passive detection to proactive enforcement of authorization, laying the groundwork for IP management of diffusion models.

\end{abstract}



\received{20 February 2007}
\received[revised]{12 March 2009}
\received[accepted]{5 June 2009}
 
\maketitle

\section{Introduction}
The rapid evolution of diffusion models—exemplified by systems such as DALL·E2 \cite{ramesh2022hierarchical}, Stable Diffusion \cite{rombach2022high}, and Instruct-Pix2Pix \cite{brooks2023instructpix2pix}—has revolutionized AI-generated imagery. Yet, this breakthrough has also precipitated widespread misuse, raising serious concerns over intellectual property protection and the unauthorized deployment of these powerful models. Traditional watermarking techniques (e.g., DWTDCT \cite{cox2007digital}, DWTDCTSVD \cite{cox2007digital}) and more recent deep learning-based methods \cite{zhu2018hidden,tancik2020stegastamp,zhang2019robust} are predominantly passive in nature. They focus on embedding latent signatures into the generated content, which can later be used to trace illicit use. However, because these approaches merely signal misuse after the fact, they fall short of actively preventing unauthorized exploitation.

\begin{figure}[t]
  \includegraphics[width=\columnwidth]{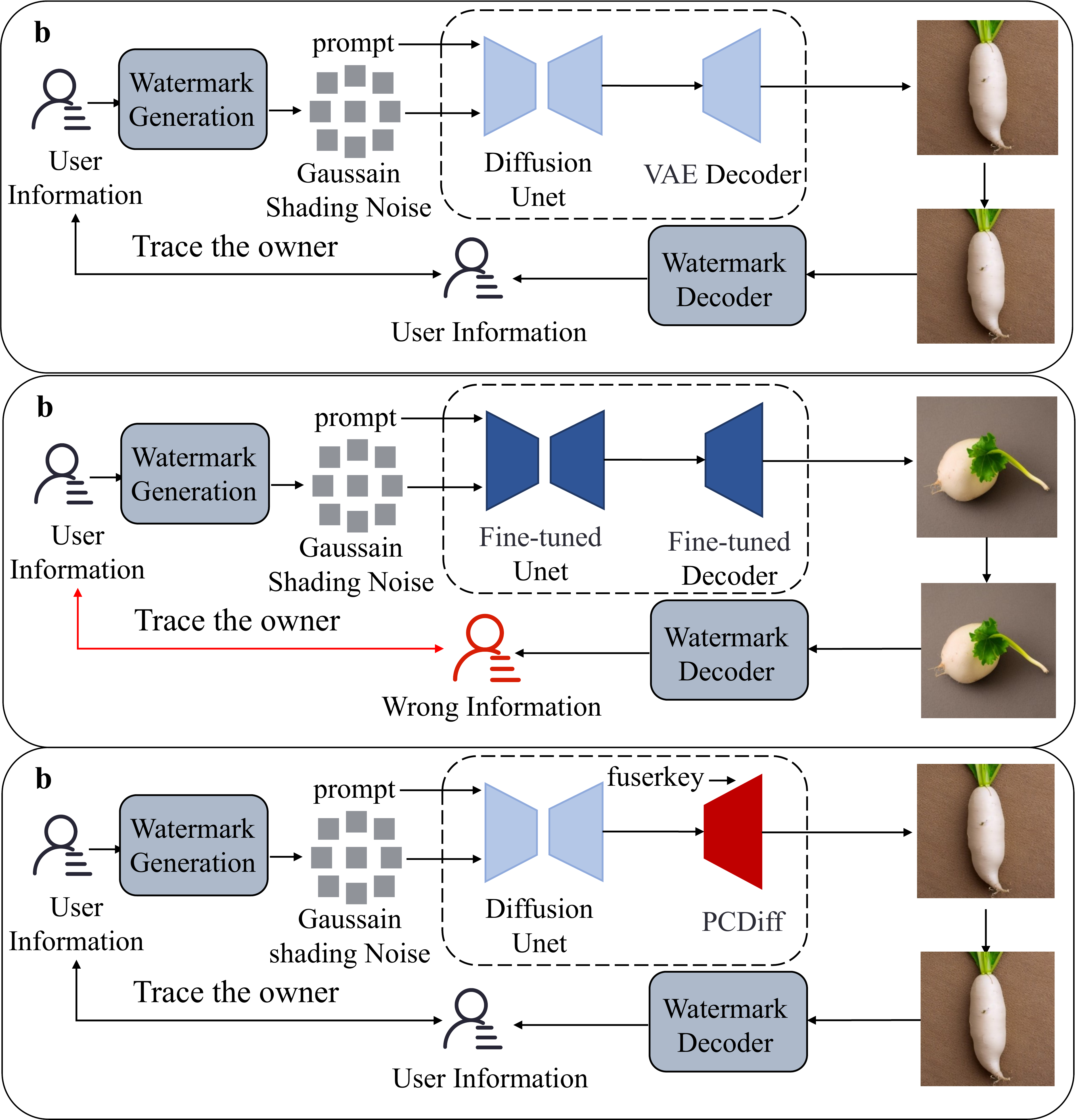}
  \caption{An illustration of the motivation behind PCDiff.}
  \Description{The figure illustrates the motivation behind PCDiff. It contains three subfigures showing different scenarios of watermark generation and extraction: (a) correct information trace with a diffusion Unet and VAE decoder, (b) wrong information trace with a fine-tuned Unet and decoder, and (c) correct information trace with a diffusion Unet and a cryptofuser. Each subfigure demonstrates the process from user information to watermark generation, embedding, and decoding.}
  \label{fig:whole}
\end{figure}

Recent research has aimed to address these vulnerabilities by incorporating watermarking directly into the diffusion process \cite{fernandez2023stable,xiong2023flexible,min2025watermark,lei2024diffusetrace,meng2024latent,ci2025ringid,peng2025intellectual}. These methods provide critical support for intellectual property protection by embedding invisible signatures into generated content, enabling post-hoc tracing and evidence collection for unauthorized use. Nevertheless, due to the inherently passive nature of watermarking techniques, they fail to proactively prevent unauthorized exploitation. Consequently, model owners seek to retain the benefits of watermarking while introducing proactive control mechanisms that can prevent misuse at the source. This dual approach aims to ensure robust protection against both current and future threats to model integrity and ownership.

Existing proactive control methods have several limitations. For example, some methods, such as DreamBooth~\cite{ruiz2023dreambooth}, Textual Inversion~\cite{gal2022image} and StyleDrop~\cite{sohn2023styledrop}, are only capable of managing specific generation conditions. In addition, techniques such as Custom Diffusion~\cite{kumari2023multi} and HyperNetworks~\cite{ruiz2024hyperdreambooth} not only require extensive parameter adjustments—which may compromise the integrity of existing watermarking mechanisms—but also rely on large-scale retraining, resulting in significant resource consumption. Moreover, most of these methods primarily focus on controlling the model's style rather than addressing ownership protection. These limitations underscore the importance of developing a novel framework that preserves watermark efficacy while enabling proactive control over model ownership.As shown in \autoref{fig:whole}, This figure compares different control methods using the Gaussian shading watermark algorithm as an example. In (a), the standard text-to-image generation process allows the embedded watermark to be successfully decoded from the generated image. In (b), existing control methods, such as fine-tuning UNet or the generative model, often distort the generated content, making watermark decoding ineffective. 

In this work, we propose PCDiff, a hierarchical authorization framework transforming a diffusion model into a credential-\linebreak dependent system. Instead of merely embedding an invisible mark, our approach integrates a proactive security mechanism directly into the generative process. By incorporating specialized fuser layers and trainable layers within the decoder, we dynamically couple cryptographic key validation—via a FuserKey with core noise prediction operations. Consequently, only users with the correct FuserKey can utilize the model's full generative capabilities; unauthorized use results in systematic output quality degradation.
Crucially, by freezing the original model structure during training, our design maintains existing watermarking functionalities intact while enforcing robust protection against misuse. This tight integration of access control and image synthesis ensures that unauthorized modifications, such as watermark removal, are ineffective. Moreover, this approach is lightweight, adding minimal overhead to the original model.

The principal contributions are summarized as follows. 
(1) We propose the first general proactive ownership control framework for diffusion models. This framework decouples active access control from the image synthesis process while preserving existing watermarking techniques, effectively safeguarding model ownership.
(2) We introduce a FuserKey-based secure generation mechanism by embedding specialized fuser and trainable layers within the model decoder. This design dynamically couples cryptographic key validation with noise prediction to degrade output quality for unauthorized access.
(3) Experimental evaluations demonstrate that our framework can effectively enforce ownership control and preserve watermark efficacy under various attack scenarios, while introducing acceptable overhead in terms of image quality and model performance.


\section{Related Work}

\subsection{Active Authorization}
Current access control mechanisms for generative AI broadly bifurcate into cryptographic watermarking and architectural interventions. Traditional watermarking approaches, such as DeepSigns~\cite{darvish2019deepsigns}, embed imperceptible markers within activation distributions to enable ownership verification, yet remain vulnerable to adaptive attacks as shown in robust removal studies~\cite{namba2019robust}. Recent textual inversion techniques~\cite{gal2022image} demonstrate concept-specific embedding manipulation, suggesting dynamic watermark embedding through latent space modifications could enhance robustness. 
In architectural control, while GAN-Control~\cite{shoshan2021gan} enables style manipulation through disentangled latents, its rigidity contrasts with Custom Diffusion's~\cite{kumari2023multi} lightweight cross-attention tuning that preserves 97\% original model capacity. The hierarchical conditional decoding in ControlAR~\cite{li2024controlar} further proves multi-level control feasibility through autoregressive token fusion. DreamBooth~\cite{ruiz2023dreambooth} introduces class-preservation loss that prevents semantic drift, a critical feature for maintaining permission policies. 
Hardware-bound solutions like DeepAttest~\cite{chakraborty2020hardware} face scalability challenges, as highlighted in diffusion control surveys~\cite{cao2024controllable}. Previous works on active control largely focus on semantic regulation, whereas our proposed PCDiff directly modulates generation quality to assert control over model ownership.

\subsection{Watermarking in Diffusion Models}
Diffusion models have shown exceptional performance in image generation tasks, and researchers have begun exploring watermark embedding during the generation process to protect copyright. Wen et al. \cite{wen2023tree} proposed tree ring watermarks, which adjust the frequency domain of latent representations to convey copyright information but are limited to single-bit watermarks. Yang et al. \cite{yang2024gaussian} introduced Gaussian Shading, mapping watermarks to standard Gaussian distributions in the latent space, supporting multi-bit watermarks but facing evasion risks. Min et al. \cite{min2025watermark} proposed \textit{WaDiff}, embedding watermarks directly into the UNet backbone for user identification and robustness. AquaLoRA \cite{feng2024aqualora} enhances flexibility through a two-stage LoRA module. Stable Signature \cite{fernandez2023stable} fine-tunes the VAE decoder to support user traceability.

Xiong et al. \cite{xiong2023flexible} developed an end-to-end watermarking method based on the ENDE architecture, allowing flexible message changes and preventing watermark bypass but requiring fine-tuning of the entire VAE decoder. Liu et al. proposed T2IW \cite{liu2023t2iw}, enforcing compatibility between semantic features and watermark signals at the pixel level for high-quality generation and robust embedding.

\section{Problem Definition}
\begin{figure*}[t]
  \centering
  \includegraphics[width=\textwidth]{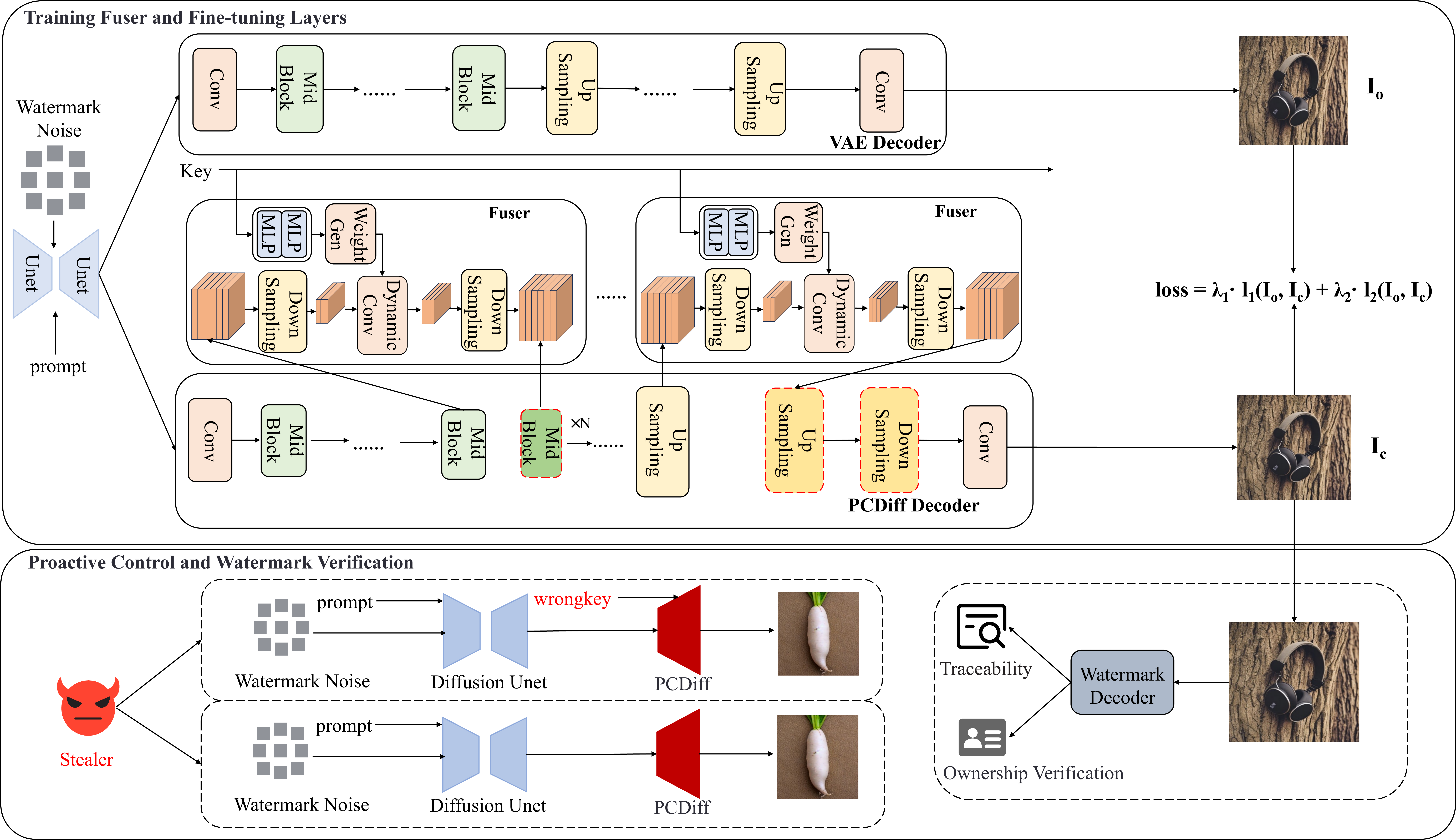}
  \caption{The training strategy and framework of proposed ownership protection method for diffusion models.}
  \Description{A schematic overview of the system architecture. The diagram illustrates the complete process flow for active control: it begins with the generation pipeline, incorporates watermark embedding and cryptographic key integration, and continues through the quality degradation steps that enforce model ownership. Key components are highlighted to show real-time authorization and verification pathways, as well as backward compatibility with legacy systems.}
  \label{fig:vae}
\end{figure*}

In this work, we address the challenge of protecting diffusion models from the unauthorized generation of high-quality images. 
Traditional watermarking methods serve as a passive post-hoc measure for verification and traceability of ownership. 
In contrast, our framework actively enforces access control by embedding a cryptographic key validation mechanism directly into the image synthesis process. 
Meanwhile, watermarking remains effective in protecting model ownership even within the PCDiff framework.


\noindent{\bf Authorized Generation.}
Within the PCDiff framework, the generation of high-quality images is explicitly gated by a valid cryptographic key, referred to as the \emph{FuserKey}. 
Let $M$ denote a diffusion model and $I$ be the output image generated with a key $K$. 
We define the image generation process by Eq. (\ref{eq:1}), where $Q(I)$ measures the quality of the generated image and $\tau$ is a predefined quality threshold. 
\begin{equation}\label{eq:1}
    I = \text{Generate}(M, K) \quad \text{with } Q(I) \geq \tau \quad \text{if } K = K_{\text{valid}}.
\end{equation}
When an invalid key (or no key) is provided, the system deliberately produces an image $I'$ with quality $Q(I') < \tau$, thereby preventing unauthorized users from obtaining high-quality outputs.

\noindent{\bf Watermarking in PCDiff.}
Watermarking techniques remain effective within the PCDiff framework for model ownership verification and provenance tracking. 
Specifically, a unique imperceptible watermark $W$ can be embedded into the generated image $I$; thus, there exists $W = \text{Extract}(I, M)$, where ownership is verified if the extracted watermark matches the registered watermark $W_{\text{owner}}$ ($W = W_{\text{owner}}$).
By embedding a model-specific identifier into $w$, traceability is enabled through $ID = \text{Decode}(W)$, which enables linking the generated image back to its source model.


\subsection{Threat Model}
We define the goal, knowledge, and capability of adversaries.

\noindent{\bf Adversary's Goal:}
The primary goal of an adversary is to bypass PCDiff’s cryptographic access control to generate high-quality images without possessing the valid FuserKey.
It implies that the adversary can obtain high-quality outputs and potentially misuse the model without detections by achieving the goal. 
The mathematical expression of the adversary aim is defined by Eq. (\ref{eq:2}), where $K' \neq K_{\text{valid}}$. 
\begin{equation}\label{eq:2}
I' = \text{Generate}(M, K') \quad \text{while  } Q(I') \geq \tau.
\end{equation}

\noindent{\bf Adversary's Knowledge:}
The adversary is assumed to have full access to the diffusion model $M$ (e.g., through theft or unauthorized distribution) and may possess in-depth knowledge of its architecture, including the embedded cryptographic mechanism.

\noindent{\bf Adversary's Capabilities:}
(1) \emph{Module Tampering:} Directly removing or altering the fuser and trainable layers that enforce the FuserKey-based validation, thereby attempting to bypass the quality degradation mechanism.
(2) \emph{Image Post-Processing:} Utilizing image enhancement or restoration techniques to recover high-quality images from outputs that have been intentionally degraded due to unauthorized access.
(3) \emph{Unauthorized Key Injection:} Attempting to generate high-quality images by supplying random or adversarially chosen FuserKeys in an effort to manipulate the model's quality degradation mechanism.


\section{Method}
\subsection{Overview of Framework}
Our proposed PCDiff framework is designed to safeguard the intellectual property of text-to-image diffusion models by actively controlling generation quality through cryptographic validation. The framework is seamlessly integrated into the model's decoder architecture via specialized fuser and Fine-tuning layers that condition the latent representations on a provided cryptographic credential, referred to as the FuserKey.

Specifically, the PCDiff operates in three main phases.

\textit{PCDiff Structure Selection}: Based on the computational resources and security requirements, the appropriate numbers of fuser and fine-tuning layers are selected.

\textit{Training Fuser and Fine-tuning Layers}: The rest of the model's architecture is frozen, while the fuser and fine-tuning layers are trained with the objective of producing images that closely resemble those generated by the original model. This approach ensures that the watermarking method remains unaffected.

\textit{Proactive Control and Watermark Verification}: Even if a model thief manages to steal the model, without the correct FuserKey, high-quality image generation is blocked. Meanwhile, a verifier can still authenticate and trace the model's ownership through the watermarks embedded in the images.



By coupling cryptographic validation with the diffusion process, PCDiff ensures that only authorized users can fully utilize the generative capabilities of the model, effectively protecting the model owner’s intellectual property.

\subsection{Training the Selected PCDiff Structure}

By integrating into the decoder of the stable diffusion model a number of fuser and fine-tuning layers—where the quantity is determined by the user's computational resources and security requirements—that mirror the structure of the other decoder layers, including the mid block and upsampling layers, PCDiff is capable of actively controlling the generation quality of the diffusion model. By freezing the remaining layers and training the newly added structures with the objective of maintaining outputs that are as consistent as possible with those of the original model, the effectiveness of both the latent space watermark and the fine-tuning watermark in the original model is preserved.

\subsubsection{Fuser Layer Design}:The fuser layer is the core component for enforcing proactive control, ensuring high-quality images are generated only when the correct FuserKey is provided. As illustrated in \autoref{fig:vae}, the fuser layer first transforms the input bit sequence (e.g., FuserKey) into a feature vector via an embedding module. This transformation maps the input into a high-dimensional space, enabling effective interaction with image features. The resulting feature vector is then processed by a dynamic convolution module, which modulates the image features based on the provided FuserKey to control the output quality.

To preserve essential image information, residual connections are integrated into the fuser layer, allowing unaltered features to bypass the convolution module and mitigate potential distortions introduced during the watermarking process. This design ensures that the watermark mechanism is tightly bound to the correct FuserKey. In the absence of the correct key, the image quality deteriorates significantly, preventing unauthorized use or tampering.



\begin{algorithm}[htbp]
\caption{Fuser Layer Forward Process}
\label{alg:fuser}
\begin{algorithmic}[1]
\Require{Image feature $F \in \mathbb{R}^{C \times H \times W}$, Fuser key $K \in \{0,1\}^d$} 
\Ensure{Modified feature $F' \in \mathbb{R}^{C \times H \times W}$}

\State $V \leftarrow \mathrm{Embedding}(K)$
\State $F_{\text{dyn}} \leftarrow \mathop{\mathrm{DynamicConv}}\nolimits{}_{W}(F, V)$
\State $F' \leftarrow F_{\text{dyn}} + F$

\Return $F'$
\end{algorithmic}
\end{algorithm}

\subsubsection{Fine-tuning Layers Design}:In addition to the fuser layers, we introduce Fine-tuning layers that adopt the same structure as the original decoder layers. These layers are inserted at multiple stages of the decoder to further enhance security and robustness. Specifically, as depicted in Figure 3, we incorporate Fine-tuning layers (highlighted in red dashed boxes) both before and after the mid-block of the decoder. In this design, the fuser layers interact with the cryptographic FuserKey (and, optionally, auxiliary Gaussian watermark noise) to dynamically modulate the latent feature representations, while the Fine-tuning layers refine the decoded features, ensuring that the intended security mechanism is tightly integrated into the output.

Furthermore, during the upsampling stage, Fine-tuning upsampling layers are introduced to improve the model's adaptability to the transformations imposed by the security mechanism. To maintain consistent output dimensions and avoid distortion, a corresponding downsampling layer is appended after the upsampling operation. This pairing preserves the overall structure of the decoder while enforcing the integration of the security functionality.

\autoref{fig:vae} highlights the seamless integration of the fuser layers and Fine-tuning layers within the VAE decoder. This structural enhancement not only degrades the quality of the generated images when an incorrect fuserkey is provided, but also ensures that any model stealer cannot easily remove the PCDiff.

\subsubsection{Training}
During the training phase, we employ two models: the original diffusion model as a high-quality reference and another incorporating the PCDiff architecture. In this configuration, the encoder and original decoder layers are frozen to preserve the effectiveness of the original watermarking algorithm, while only the newly introduced fuser and Fine-tuning layers are optimized.

For the training process, we employ two loss functions: Mean Absolute Error (MAE) and the Learned Perceptual Image Patch Similarity (LPIPS) \cite{zhang2018unreasonable}. These loss functions ensure that the images generated by the enhanced model closely approximate those produced by the original model. The overall loss function is defined as follows:
\begin{equation}
L = \lambda_1 \cdot l_{MAE} + \lambda_2 \cdot l_{lpips}
\end{equation}
where $\lambda_1$ and $\lambda_2$ are both set to 1, and the learning rate is configured to $ 1\mathrm{e}{-5} $

\subsection{Proactive Control and Watermark Verification}
As shown in \autoref{fig:vae}, When an attacker steals a diffusion model and attempts to use it for image generation or claim ownership of the model, they will find that, under the protection of PCDiff, the model fails to generate high-quality images. This effectively achieves proactive protection of the model's ownership. Meanwhile, if the attacker attempts to claim ownership of the images generated by the original model owner, PCDiff ensures the effectiveness of watermarking. The watermarking algorithm can successfully extract the watermark from the generated images, enabling ownership verification and provenance tracing.

When an attacker obtains the diffusion model protected by PCDiff, due to the lack of knowledge of the correct FuserKey (which is set to 128 bits in our design), the probability of successfully guessing the correct FuserKey through brute-force search is approximately \(1/2^{128}\), which is computationally infeasible. Therefore, the attacker is highly likely to attempt generating images using an incorrect FuserKey or even removing the fuser to generate images directly. As shown in \autoref{fig:vae}, the attacker fails to generate normal high-quality images, rendering the stolen model practically unusable.

Assuming the attacker tries to remove the fine-tuning layers embedded in the decoder of the diffusion model, it remains extremely challenging. Since the fine-tuning layers are designed to have the same structure as the corresponding layers in the original decoder, the attacker cannot easily distinguish them. The only feasible approach is to exhaustively test different combinations of layer removals.

For a theoretical analysis of the time required for cracking, we adopt the default settings from the official Stable Diffusion code~\cite{rombach2021highresolution}, where the original model comprises two mid blocks and performs three upsampling operations. Assuming that \( m \) denotes the number of newly added mid blocks and \( n \) denotes the number of newly added upsampling layers, the total number of possible combinations is given by
\begin{equation}
C_{m+2}^{2} + (n+1)^3.
\end{equation}

Consequently, the theoretical time required for cracking is computed as
\begin{equation}
T_{crack} = t_{test} \times \left[ C_{m+2}^{2} + (n+1)^3 \right]
\end{equation}

Therefore, as the number of fine-tuning layers increases, the cracking time grows exponentially, and the computational resources required by the attacker also increase exponentially. In this way, PCDiff effectively achieves proactive protection of the model's ownership.

If the attacker fails to compromise the model itself, they may instead attempt to steal the images generated by the model owner, falsely claim ownership of these images, and disseminate them without restriction. However, due to the design of PCDiff, the original model watermarking algorithm remains fully effective.

When the model owner or any third party obtains the images claimed by the attacker, they can apply the corresponding watermark decoding algorithm to extract the embedded watermark. Based on the decoded watermark information, the true ownership of the image can be verified, and the origin of the disseminated images can be traced back to their legitimate creator.

In this way, the original functionality of the model watermarking algorithm remains intact and can still be used effectively for ownership verification and provenance tracing. 

\section{Experiment}

\subsection{Experimental Setup}

\subsubsection{Model and Training Configuration}  
We used Stable Diffusion (v2.1) and a dataset of 1,618 diverse text-to-image prompts to evaluate generation quality. All images were resized to \(512 \times 512\). The configuration with four fine-tuned mid blocks and one fine-tuned up/downsampling layer was trained for 3 epochs on two NVIDIA 3090 GPUs (24GB VRAM each). For the **86**, **204**, and **208** variants, training ran for approximately 24 hours on a single NVIDIA L20 GPU (48GB). The AdamW optimizer~\cite{loshchilov2017decoupled} was used with an initial learning rate of \(5 \times 10^{-5}\), along with gradient scaling and clipping.

\subsubsection{Evaluation Metrics and Protocol}
Image quality was assessed using PSNR~\cite{wang2004image}, SSIM, and FID~\cite{heusel2017gans} (higher PSNR/SSIM and lower FID indicate better quality). Watermarking performance was measured by the average bit accuracy of the extracted watermark. Model efficiency was evaluated by additional parameter count, average generation time, and GPU memory usage during inference.

\subsection{Proactive Control Evaluation}
To assess the impact of PCDiff on image quality and its robustness against model theft, we evaluated three PCDiff configurations. The first variant, labeled “86”, incorporates 6 finetuned mid blocks and 5 finetuned upsampling/downsampling layers (the original model includes 2 mid blocks and 1 upsampling layer). The second variant (“204”) employs 18 finetuned mid blocks and 3 finetuned upsampling/downsampling layers, and the third (“208”) uses 18 finetuned mid blocks and 7 finetuned upsampling/downsampling layers. For each configuration, 500 images were generated, and the PSNR, SSIM, and FID scores were computed using the original model as the benchmark.

In an attack scenario where an adversary gains access to both the model and its configuration files, we evaluated two conditions: (1) the fuser is removed, allowing direct image generation (``no fuser''), and (2) the fuser is retained but operated with a random key (``wrong key'') instead of the correct key. For both conditions, 500 images were generated per configuration, and the resulting average PSNR, SSIM, and FID metrics were compared with those of the original model.
\begin{table*}[t]
  \centering
  \caption{Performance comparison of different CryptoFuser configurations.}
  \setlength{\tabcolsep}{8pt} 
  \begin{tabular}{l 
    R{1.5cm}@{\hspace{5mm}}R{1.2cm}
    R{1.5cm}@{\hspace{5mm}}R{1.2cm}
    R{1.5cm}@{\hspace{5mm}}R{1.2cm}
    R{1.5cm}@{\hspace{5mm}}R{1.2cm}}
    \toprule
    \textbf{Metric} & \multicolumn{2}{c}{\textbf{ori (Base)}} & \multicolumn{2}{c}{\textbf{8 6}} & \multicolumn{2}{c}{\textbf{20 4}} & \multicolumn{2}{c}{\textbf{20 8}} \\
    \cmidrule(lr){2-3} \cmidrule(lr){4-5} \cmidrule(lr){6-7} \cmidrule(lr){8-9}
    \textbf{Model Params (B)} 
      & 1.30 & (0\%) 
      & 1.54 & (18.46\%) 
      & 1.54 & (18.46\%) 
      & 1.54 & (18.46\%) \\
      
    \textbf{Model Size (MB)}
      & 4972.83 & (0\%) 
      & 5877.21 & (18.18\%) 
      & 5877.21 & (18.18\%) 
      & 5877.21 & (18.18\%) \\
      
    \textbf{Total Time (s)}
      & 661.29 & (0\%) 
      & 730.09 & (10.40\%) 
      & 1051.27 & (58.98\%)
      & 1053.85 & (59.36\%) \\
      
    \textbf{Avg/Batch (s)}
      & 6.61 & (0\%) 
      & 7.30 & (10.41\%) 
      & 10.51 & (58.99\%)
      & 10.54 & (59.40\%) \\
      
    \textbf{Fastest Batch (s)}
      & 6.44 & (0\%) 
      & 7.01 & (8.81\%) 
      & 6.82 & (5.86\%)
      & 7.03 & (9.11\%) \\
      
    \textbf{Slowest Batch (s)} 
      & 15.83 & (0\%) 
      & 16.08 & (1.58\%) 
      & 16.42 & (3.73\%)
      & 20.32 & (28.35\%) \\
      
    \textbf{Peak Mem (GB)}
      & 8.09 & (0\%) 
      & 9.99 & (23.43\%) 
      & 10.10 & (24.82\%)
      & 10.10 & (24.82\%) \\
      
    \textbf{Avg Mem (GB)}
      & 8.01 & (0\%) 
      & 9.91 & (23.72\%) 
      & 10.02 & (25.09\%)
      & 10.02 & (25.09\%) \\
    \bottomrule
  \end{tabular}
  \label{tab: Performance Evaluation}
\end{table*}
As shown in Table~\ref{tab:security}, under normal conditions the image quality remains largely preserved (with only minor reductions in PSNR, SSIM, and FID). However, when either the fuser is removed or an incorrect key is used, quality deteriorates substantially. These results indicate that, while the integration of \texttt{PCDiff} incurs only minimal quality loss under normal operation, its absence or misuse leads to significant degradation, effectively deterring unauthorized use.

\begin{table}[t]
  \centering
  \caption{Quantitative evaluation of image quality under different structures and conditions}  
  \begin{tabular}{ccccc}
    \toprule
    \textbf{Structure} & \textbf{Condition} & \textbf{PSNR} & \textbf{SSIM} & \textbf{FID} \\
    \midrule
    86  & ori        & 28.0088 & 0.9076 & 8.5121 \\
    86  & wrong key  & 24.6234 & 0.8664 & 36.2917 \\
    86  & no fuser   & 22.1754 & 0.8340 & 36.6722 \\
    \midrule
    204 & ori        & 30.1266 & 0.9095 & 8.5375 \\
    204 & wrong key  & 25.3485 & 0.8533 & 34.8126 \\
    204 & no fuser   & 21.4896 & 0.8221 & 39.3671 \\
    \midrule
    208 & ori        & 28.4268 & 0.8778 & 11.2761 \\
    208 & wrong key  & 22.2775 & 0.8134 & 44.2957 \\
    208 & no fuser   & 17.2655 & 0.7316 & 63.8445 \\
    \bottomrule
  \end{tabular}
  \label{tab:security}
\end{table}

\begin{figure}[htbp]
  \centering
  \begin{subfigure}{0.3\linewidth}
    \centering
    \label{}
    \includegraphics[width=\linewidth]{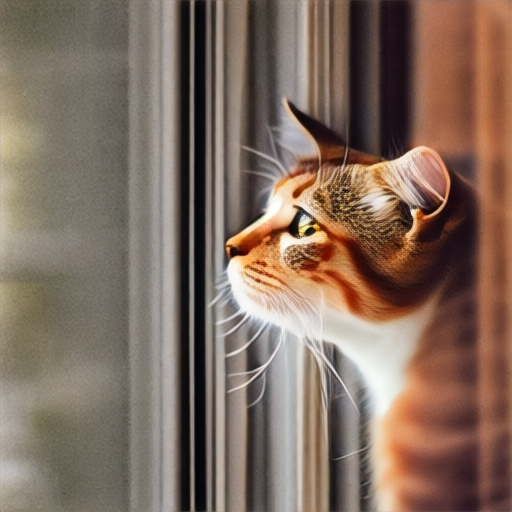}
    \caption{86, ori\\PSNR:28.69dB SSIM:0.9444}
    \label{fig:86_ori}
  \end{subfigure}
  \hspace{0.02\linewidth}
  \begin{subfigure}{0.3\linewidth}
    \centering
    \includegraphics[width=\linewidth]{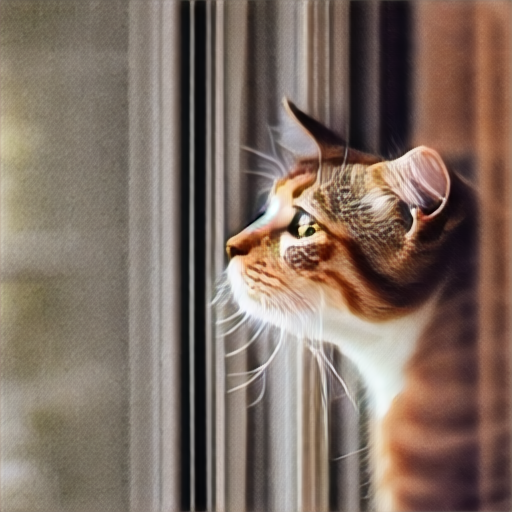}
    \caption{86, wrong key\\PSNR:26.16dB SSIM:0.9217}
    \label{fig:86_wrong_key}
  \end{subfigure}
  \hspace{0.02\linewidth}
  \begin{subfigure}{0.3\linewidth}
    \centering
    \includegraphics[width=\linewidth]{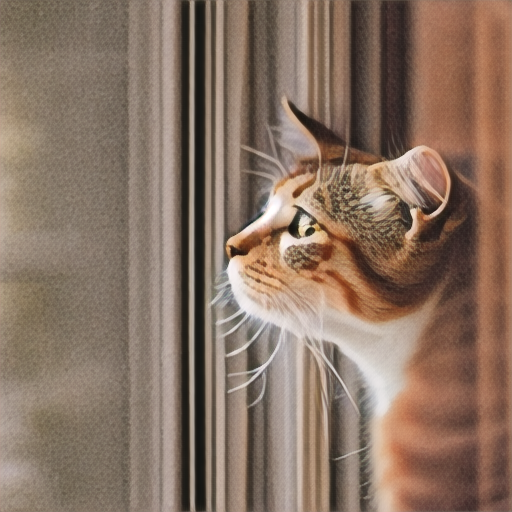}
    \caption{86, no fuser\\PSNR:24.41dB SSIM:0.8992}
    \label{fig:86_no_fuser}
  \end{subfigure}

  \vspace{0.8em}

  \begin{subfigure}{0.3\linewidth}
    \centering
    \includegraphics[width=\linewidth]{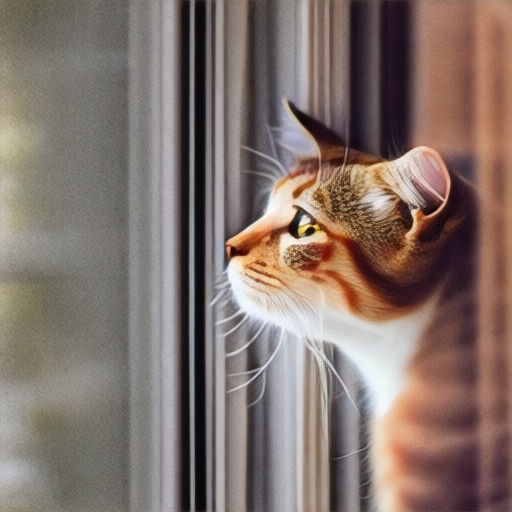}
    \caption{204, ori\\PSNR:32.31dB SSIM:0.9463}
    \label{fig:204_ori}
  \end{subfigure}
  \hspace{0.02\linewidth}
  \begin{subfigure}{0.3\linewidth}
    \centering
    \includegraphics[width=\linewidth]{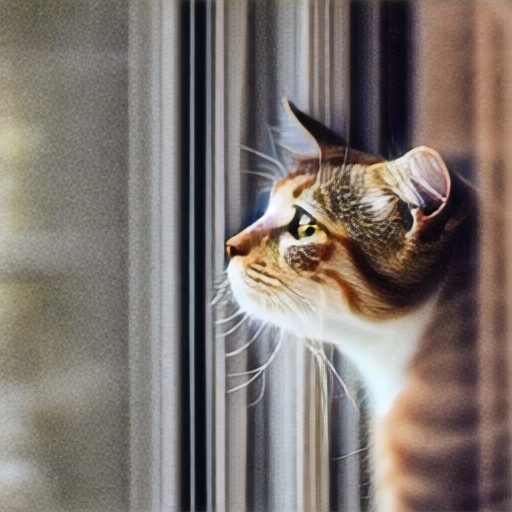}
    \caption{204, wrong key\\PSNR:27.25dB SSIM:0.9086}
    \label{fig:204_wrong_key}
  \end{subfigure}
  \hspace{0.02\linewidth}
  \begin{subfigure}{0.3\linewidth}
    \centering
    \includegraphics[width=\linewidth]{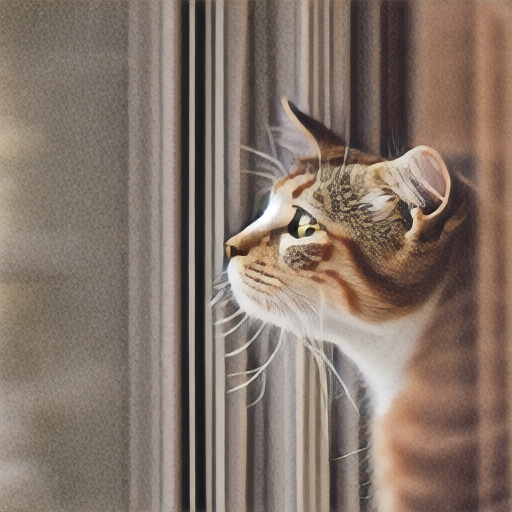}
    \caption{204, no fuser\\PSNR:23.93dB SSIM:0.8895}
    \label{fig:204_no_fuser}
  \end{subfigure}

  \vspace{0.8em}

  \begin{subfigure}{0.3\linewidth}
    \centering
    \includegraphics[width=\linewidth]{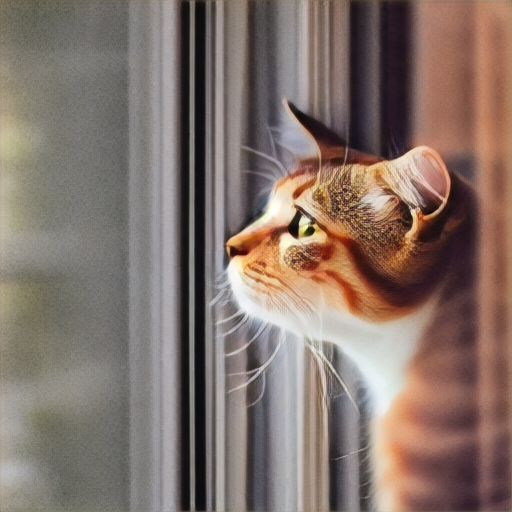}
    \caption{208, ori\\PSNR:30.51dB SSIM:0.9284}
    \label{fig:208_ori}
  \end{subfigure}
  \hspace{0.02\linewidth}
  \begin{subfigure}{0.3\linewidth}
    \centering
    \includegraphics[width=\linewidth]{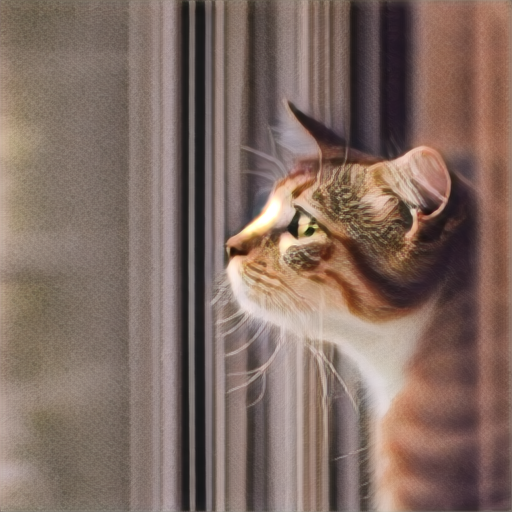}
    \caption{208, wrong key\\PSNR:23.92dB SSIM:0.8844}
    \label{fig:208_wrong_key}
  \end{subfigure}
  \hspace{0.02\linewidth}
  \begin{subfigure}{0.3\linewidth}
    \centering
    \includegraphics[width=\linewidth]{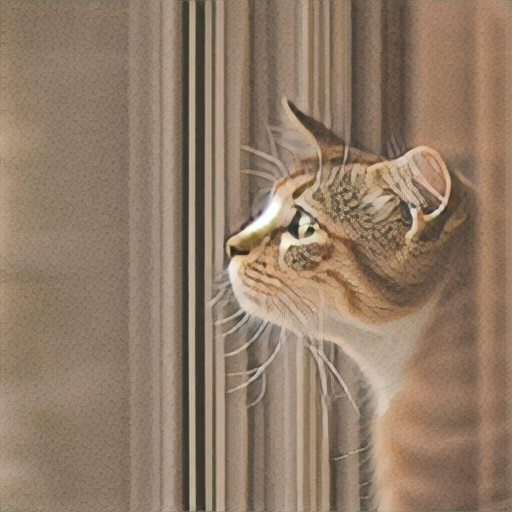}
    \caption{208, no fuser\\PSNR:18.64dB SSIM:0.8041}
    \label{fig:208_no_fuser}
  \end{subfigure}

  \caption{Visual comparison of images generated under different structures and conditions}
  \Description{Comparison of generated images using different keys (correct vs.\ wrong) or no fuser module, along with the resulting PSNR and SSIM metrics. The first, second, and third rows respectively show samples 86, 204, and 208 under various configurations. Lower PSNR and SSIM generally indicate greater degradation or distortion, highlighting how the use of the correct key and fuser preserves higher image fidelity.}
  \label{fig:all_structures_three_images}
\end{figure}

\autoref{fig:all_structures_three_images} presents visual comparisons for images generated with the prompt “\emph{a cat looking out of a window}” under three conditions: normal operation (using the correct fuser key), using a random key, and with the fuser removed. In the normal condition, images generated by all three \texttt{PCDiff} configurations (86, 204, and 208) exhibit high visual quality and consistency, serving as the baseline. When a random key is applied, the images across all configurations retain some resemblance to the normal outputs but display noticeable artifacts and color distortions. The degradation is more severe when the fuser is removed entirely, with the 208 configuration showing particularly pronounced visual inconsistencies. These observations, together with the quantitative metrics in Table~\ref{tab:security}, provide compelling evidence of the security mechanism's effectiveness in hampering unauthorized image generation.

\subsection{Impact on Model Performance}
To evaluate the impact of PCDiff on model performance, we compared three configurations (86, 204, and 208) against the original diffusion model. \autoref{tab: Performance Evaluation} summarizes key metrics—model parameters, model size, total generation time, average, fastest, and slowest batch times, as well as peak and average GPU memory usage—with relative differences shown in parentheses. The PCDiff models exhibit an approximate 18\% increase in both parameters and size. Although generation times for the 204 and 208 configurations increase by nearly 59\% and GPU memory usage rises by around 24–\%, the fastest batch times increase only modestly (5.86–9.11\%), while the slowest batch time for the 208 configuration increases by 28.35\%. Overall, these results show that the additional computational overhead is manageable, making the trade-off acceptable given the benefits in model performance and security.

\subsection{Evaluation of Attacks on Proactive Control}
\subsubsection{Brute-Force Attack Analysis}
In this section, we assume that the attacker has obtained both the model and its corresponding configuration file and, without knowledge of the fuser key, attempts to brute-force the model architecture. 

To address this vulnerability, we conducted experiments with different configurations while keeping the number of fusers constant. The tested structures include a configuration with six fine-tuned mid blocks and one fine-tuned upsampling/downsampling layer, as well as the previously introduced "86", "204", and "208" variants. By evaluating these configurations, we aim to analyze the impact of increasing the number of fine-tuned layers on the model's robustness against brute-force attacks.

\begin{figure}[t]
  \centering
  \includegraphics[width=\columnwidth]{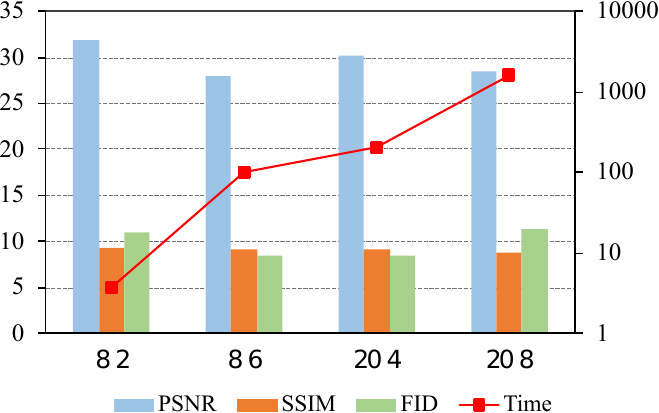}
  \caption{Relationship between image quality and theoretical cracking time for various PCDiff configurations, as measured by PSNR, SSIM, and FID—with estimated cracking times plotted on a base‑10 logarithmic scale. Note that the SSIM values are multiplied by 10 in the figure.}
  \Description{Illustration of performance metrics across four samples (82, 86, 204, 208). The bar groups display PSNR (in blue), SSIM (in orange),  and FID (in green) on the left vertical axis, while the red line indicates processing time on the right vertical axis. Higher PSNR and SSIM typically indicate better image fidelity, whereas lower FID values suggest improved generative quality. The time measurement reflects computational overhead across these configurations.}
  \label{fig:atktime}
\end{figure}

Due to the insufficient memory of a 3090 GPU, we switched to an L20 GPU with 48GB of VRAM, and the training of each PCDiff structure required an average duration of 24 hours. We evaluated the performance of these structures using PSNR, SSIM, and FID metrics, and the experimental results are presented in \autoref{fig:atktime}. The results show that, although image quality degrades to a certain degree as the number of structural layers increases, the theoretical brute-force time increases exponentially.

\begin{figure}[t]
  \centering
  \begin{subfigure}[t]{0.32\columnwidth}
    \centering
    \includegraphics[width=\linewidth]{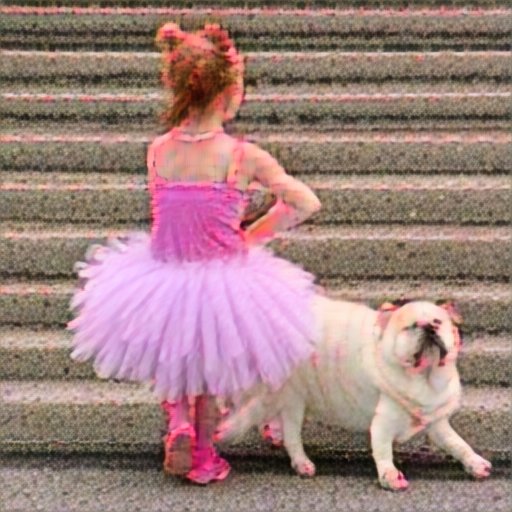}
    \caption{}
    \label{fig:arbitrary_midblock}
  \end{subfigure}
  \hfill
  \begin{subfigure}[t]{0.32\columnwidth}
    \centering
    \includegraphics[width=\linewidth]{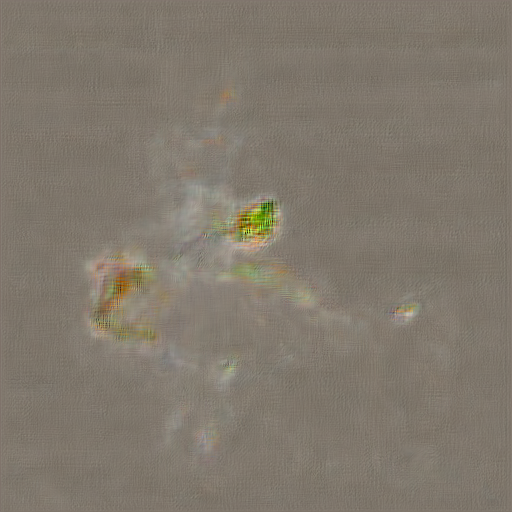}
    \caption{}
    \label{fig:arbitrary_upsampling}
  \end{subfigure}
  \hfill
  \begin{subfigure}[t]{0.32\columnwidth}
    \centering
    \includegraphics[width=\linewidth]{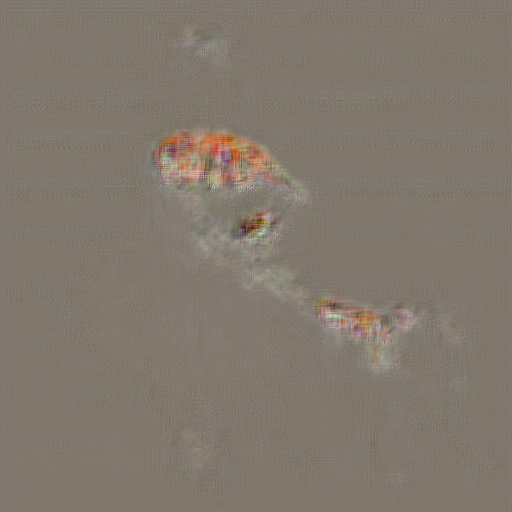}
    \caption{}
    \label{fig:arbitrary_midblock_upsampling}
  \end{subfigure}
  \vskip\baselineskip
  \begin{subfigure}[t]{0.32\columnwidth}
    \centering
    \includegraphics[width=\linewidth]{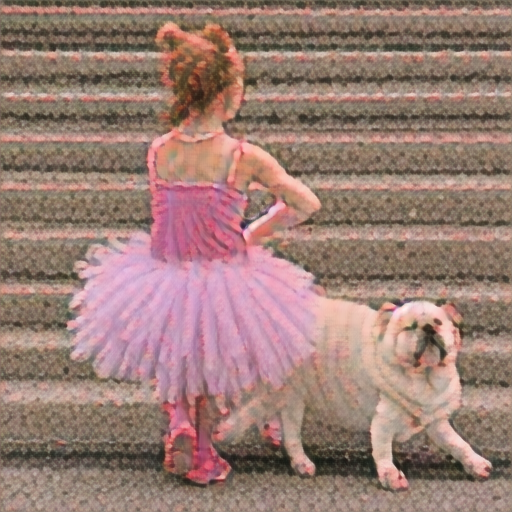}
    \caption{}
    \label{fig:removed_midblock}
  \end{subfigure}
  \hfill
  \begin{subfigure}[t]{0.32\columnwidth}
    \centering
    \includegraphics[width=\linewidth]{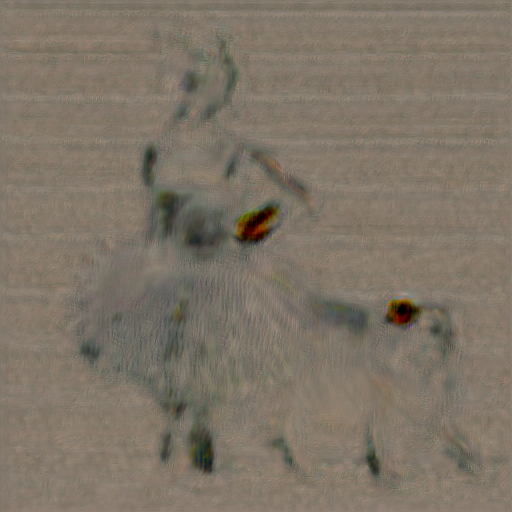}
    \caption{}
    \label{fig:removed_upsampling}
  \end{subfigure}
  \hfill
  \begin{subfigure}[t]{0.32\columnwidth}
    \centering
    \includegraphics[width=\linewidth]{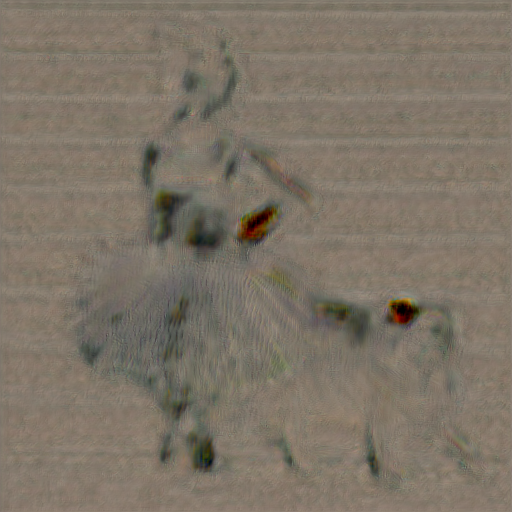}
    \caption{}
    \label{fig:removed_midblock_upsampling}
  \end{subfigure}
  \caption{Visual comparison: First row shows images from a randomly selected mid block (a), a randomly selected upsampling layer (b), and both (c). Second row shows images with some mid blocks removed (d), some upsampling layers removed (e), and both removed (f).}
  \Description{The figure illustrates a visual comparison of images processed under different conditions. The first row shows images from (a) a randomly selected mid block, (b) a randomly selected sampling layer, and (c) both combined. The second row shows images with (d) some mid blocks removed, (e) some upsampling layers removed, and (f) both removed.}
  \label{fig:three_images}
\end{figure}
\begin{table*}[t]
    \centering
    \caption{Evaluation of method robustness under nine diverse image attacks is presented. The table reports quality metrics (PSNR, SSIM, and FID); for the Treering method, performance is measured by area under the curve (AUC), while bit accuracy is reported for other methods under each attack. The attacks include: JPEG compression (QF = 75), 50\% area random crop, 30\% area random drop, Gaussian blur ($\sigma = 2$), median filtering ($k = 5$), Gaussian noise ($\mu = 0$, $\sigma = 0.1$), salt and pepper noise ($p = 0.1$), 50\% resizing, and brightness adjustment (factor = 2).}
    \begin{tabular}{lccccccccccccc}
        \toprule
        \multirow{2}{*}{\textbf{Method}} & \multicolumn{3}{c}{\textbf{Metrics}} & \multicolumn{10}{c}{\textbf{AUC/Bit Accuracy $\uparrow$}} \\
        \cmidrule(lr){2-4} \cmidrule(lr){5-14}
        & \textbf{PSNR $\uparrow$} & \textbf{SSIM $\uparrow$} & \textbf{FID $\downarrow$} 
        & \textbf{None} & \textbf{JPEG} & \textbf{Crop} & \textbf{Drop} & \textbf{GBlur} & \textbf{MFilter} & \textbf{GNoise} & \textbf{SPNoise} & \textbf{Resize} & \textbf{Bright} \\
        \midrule
        Treering& - & - & - & 1.00 & 1.00 & 0.82 & 0.98 & 1.00 & 1.00 & 1.00 & 0.99 & 1.00 & 0.99 \\
        Cf+Treeing& - & - & - & 1.00 & 1.00 & 0.73 & 0.98 & 1.00 & 1.00 & 1.00 & 0.98 & 1.00 & 0.99  \\
        Shading & - & - & - & 1.00 & 0.99 & 0.50 & 0.81 & 0.99 & 0.99 & 0.99 & 0.81 & 0.99 & 0.99 \\
        Cf+Shading & 32.68 & 0.93 & 5.05 & 1.00 & 0.99 & 0.50 & 0.81 & 0.99 & 0.99 & 0.99 & 0.81 & 0.99 & 0.99 \\
        Signature  & 33.00 & 0.89 & 3.2 & 0.98 & 0.88 & 0.92 & 0.89 & 0.55 & 0.72 & 0.97 & 0.66 & 0.88 & 0.93 \\
        Cf+Sign & 31.33 & 0.92 & 5.84 & 0.90 & 0.78 & 0.80 & 0.63 & 0.52 & 0.66 & 0.90 & 0.62 & 0.76 & 0.83 \\
        \bottomrule
    \end{tabular}
    \label{tab:full_robustness}
\end{table*}
\subsubsection{Partial Layer Removal Attack}
To prevent an attacker from bypassing protection by targeting only a few specific layers, we conducted experiments with the PCDiff configuration set to 86. In one scenario, we removed all but two randomly selected mid blocks, one randomly selected upsampling layer, or a combination of two mid blocks and one upsampling layer. In another scenario, we randomly removed a small number of mid blocks, upsampling layers, or both. As shown in \autoref{fig:three_images}, removing either mid blocks or upsampling layers leads to a substantial deterioration in image quality, effectively preventing the generation of high-quality images by attackers.

\subsubsection{Image Inpainting Attack}
We tested two restoration approaches on degraded images obtained under the attack scenarios: one using Real-ESRGAN~\cite{wang2021realesrgan} and the other using stable diffusion2 inpainting~\cite{Rombach_2022_CVPR} with the prompt “A highly detailed and realistic restoration that preserves the original content and structure of the image, only enhancing fine details.” For each configuration, 500 images were tested across three PCDiff structures (86, 204, and 208) under the two conditions ("wrong key" and "no fuser").

\autoref{tab:restoration_results} presents the restoration performance in terms of PSNR, SSIM, and FID. Analysis of the results reveals that neither restoration method is able to recover the image quality to the level of the original model. In both methods, the "wrong key" condition consistently yields better metrics than the "no fuser" condition; however, even in the best case, the restored images remain significantly inferior. This suggests that the security mechanism imposed by the fuser is robust against attempts to repair the degradation caused by unauthorized usage.
\begin{table}[htbp]
  \centering
  \caption{Restoration Results Using Real-ESRGAN and Stable-Diffusion-2-Inpainting}
  \setlength{\tabcolsep}{6pt}
  \resizebox{\columnwidth}{!}{
  \begin{tabular}{llccc}
    \toprule
    \textbf{Method} & \textbf{Structure / Condition} & \textbf{PSNR} & \textbf{SSIM} & \textbf{FID} \\
    \midrule
    \multirow{6}{*}{Real-ESRGAN} 
      & 86 / wrong key  & 24.3623 & 0.8574 & 34.2339 \\
      & 204 / wrong key  & 24.0680 & 0.8335 & 34.1290 \\
      & 208 / wrong key  & 22.3883 & 0.8097 & 42.2528 \\
      & 86 / no fuser   & 22.5241 & 0.8322 & 32.3640 \\
      & 204 / no fuser   & 21.6659 & 0.8271 & 35.4066 \\
      & 208 / no fuser   & 17.4809 & 0.7428 & 57.7411 \\
    \midrule
    \multirow{6}{*}{SD2Inpainting} 
      & 86 / wrong key  & 19.6358 & 0.6156 & 75.3901 \\
      & 204 / wrong key  & 19.5928 & 0.5852 & 78.4888 \\
      & 208 / wrong key  & 18.6782 & 0.5700 & 89.8917 \\
      & 86 / no fuser   & 18.1874 & 0.5467 & 90.8714 \\
      & 204 / no fuser   & 17.6490 & 0.5279 & 94.7671 \\
      & 208 / no fuser   & 15.6167 & 0.4893 & 114.3962 \\
    \bottomrule
  \end{tabular}
  }
  \label{tab:restoration_results}
\end{table}

\subsection{Watermark Effectiveness Evaluation}\label{4.2}

In this section, we analyze the compatibility of the PCDiff framework with various watermarking methods. To this end, we conducted experiments using a single model configuration—a diffusion model augmented with four fine-tuned mid blocks and one upsampling/downsampling layer. 

\subsubsection{Compatibility Effectiveness Test}
To assess the compatibility of PCDiff with watermarking methods, we conducted experiments on three approaches: Treering, Gaussian shading, and Stable Signature. Table~\ref{tab:full_robustness} reports image quality metrics (PSNR, SSIM, FID) alongside robustness measurements—TPR for Treering and bit accuracy for both Gaussian shading and Stable Signature—under ten post-processing attacks, including JPEG compression, random cropping, random drop, Gaussian blur, median filtering, Gaussian noise, salt–pepper noise, resizing, and brightness adjustment. Although integrating PCDiff results in minor variations in image quality and watermark performance (for instance, Stable Signature shows a slight decrease in PSNR and a moderate increase in FID), the overall performance remains within acceptable limits. These findings demonstrate that PCDiff can be integrated with different watermarking methods without causing significant degradation in either image quality or watermark robustness.
\begin{figure}[htbp]
  \centering
  \begin{subfigure}{0.48\linewidth}
    \centering
    \includegraphics[width=\linewidth]{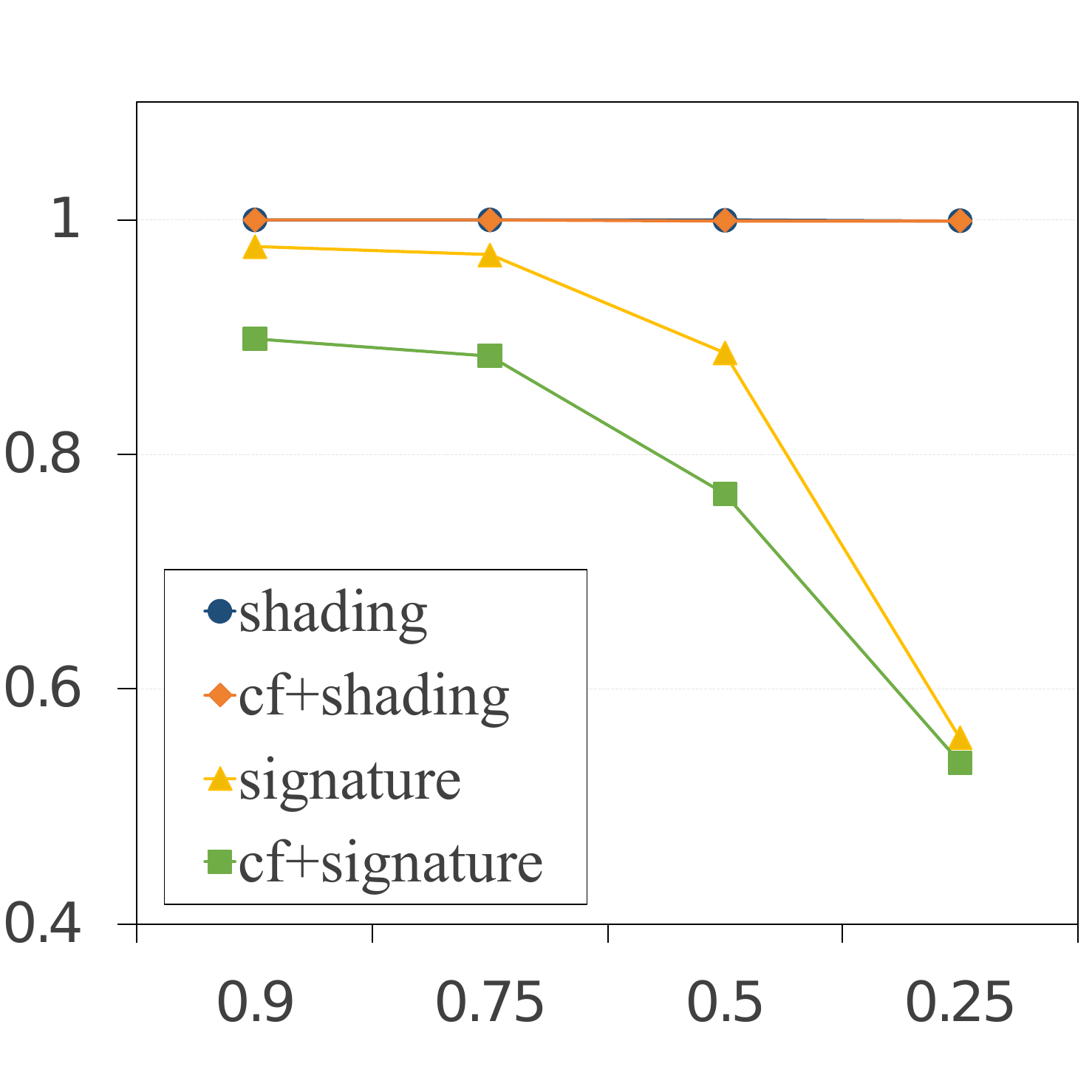}
    \caption{Resize (RE)}
    \label{fig:chart_RC}
  \end{subfigure}
  \hfill
  \begin{subfigure}{0.48\linewidth}
    \centering
    \includegraphics[width=\linewidth]{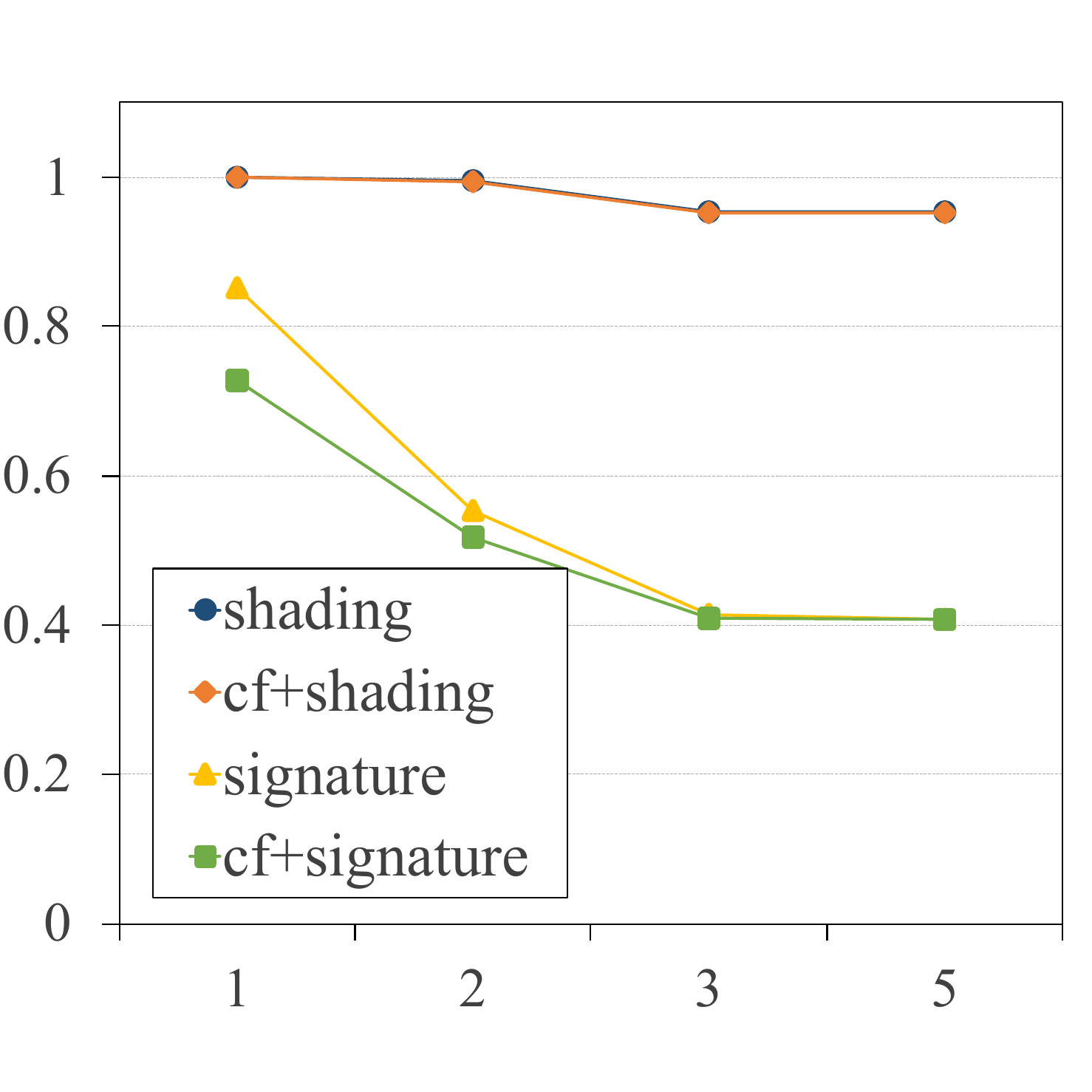}
    \caption{Gaussian Blur (GB)}
    \label{fig:chart_GB}
  \end{subfigure}
  
  \vspace{1em}
  
  \begin{subfigure}{0.48\linewidth}
    \centering
    \includegraphics[width=\linewidth]{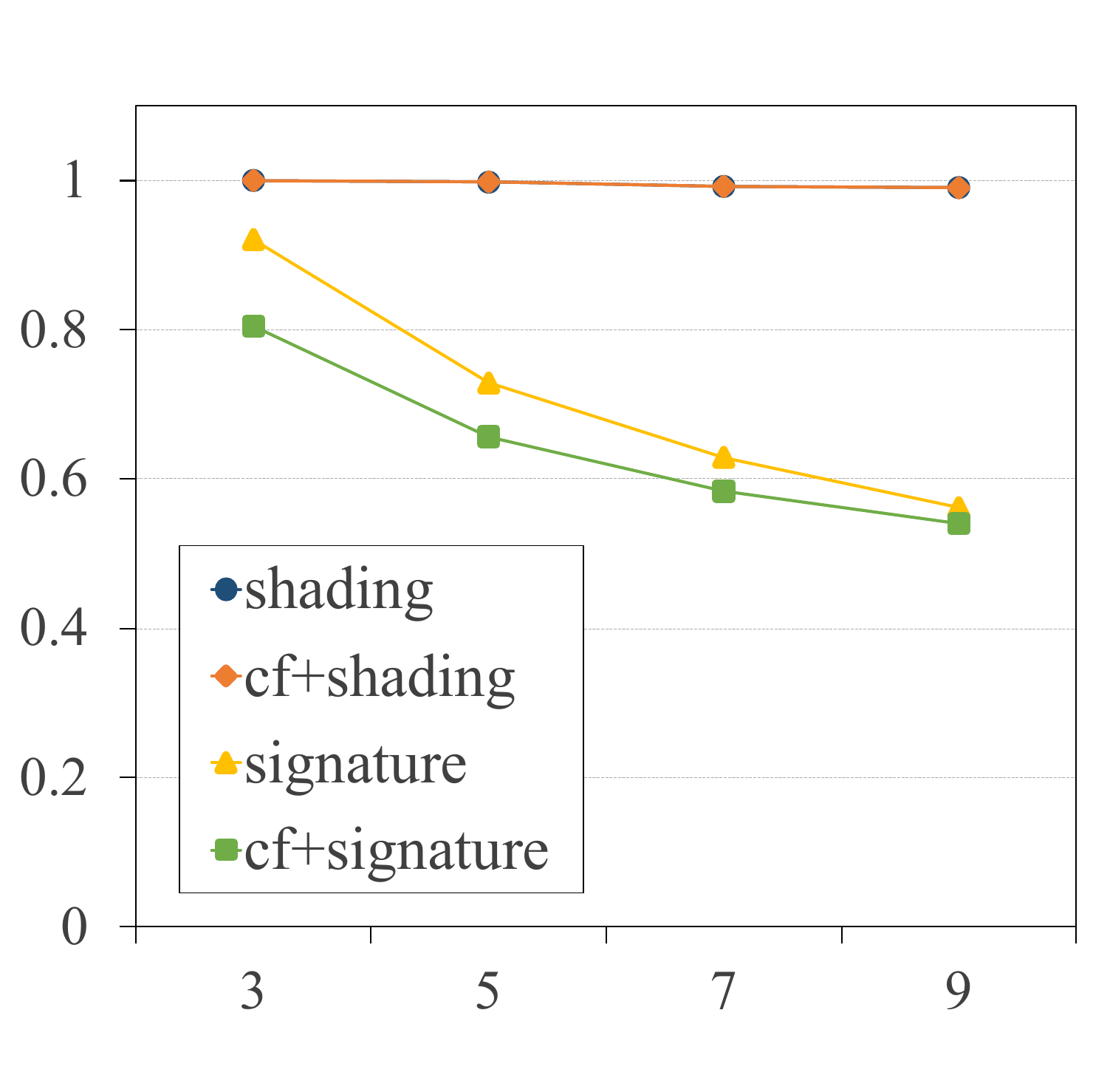}
    \caption{Median Filter (MF)}
    \label{fig:chart_MF}
  \end{subfigure}
  \hfill
  \begin{subfigure}{0.48\linewidth}
    \centering
    \includegraphics[width=\linewidth]{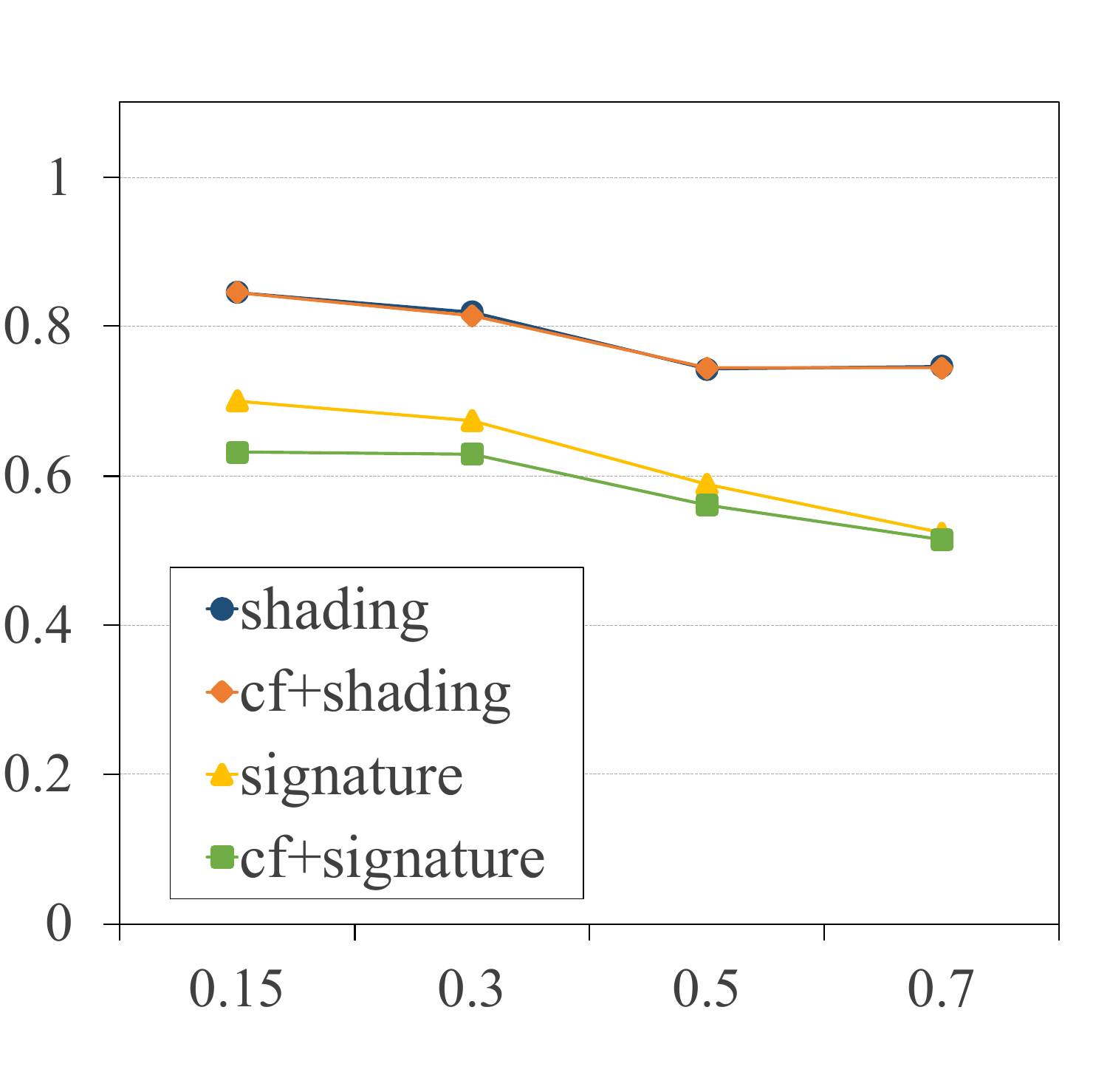}
    \caption{Random Drop (RD)}
    \label{fig:chart_RD}
  \end{subfigure}
  
  \caption{Watermark extraction accuracy under varying attack intensities for RE, GB, MF, and RD.}
  \Description{A figure showing watermark extraction accuracy under varying attack intensities. The figure contains four subfigures: Resize (RE), Gaussian Blur (GB), Median Filter (MF), and Random Drop (RD). Each subfigure illustrates the performance metrics of the watermarking technique under different attack conditions.}
  \label{fig:robustness_attack_4}
\end{figure}

\subsubsection{Robustness under Varying Attack Intensities}
In addition to evaluating robustness under fixed attack settings, we also conducted experiments across different attack intensities to assess the stability of our framework. Our results indicate that the reduction in watermark extraction accuracy introduced by the PCDiff framework remains confined within a fixed range regardless of the attack strength. Even as the severity of attacks such as JPEG compression, resizing, or brightness adjustment increases, the degradation in robustness does not exhibit any dramatic decline. Instead, the performance drop is stable and predictable, ensuring that the watermark extraction process remains reliable under a variety of conditions. Notably, while experiments were performed on nine different types of attacks, for brevity the paper only presents detailed results for four commonly encountered attacks, with the overall conclusions remaining unchanged. This stability confirms that our framework is well-suited to adapt to different attack intensities, thereby maintaining secure and accurate watermark retrieval even when confronted with harsher image processing distortions.

\section{Conclusion}

In this work, we propose PCDiff, an IP protection framework for text-to-image diffusion models. Unlike traditional watermarking, PCDiff provides proactive control over model ownership directly. The framework employs a fuser module and trainable layers to ensure that only authorized users with valid credentials can generate high-quality images. Unauthorized attempts lead to a controlled degradation in output quality. Our approach also preserves the effectiveness of existing watermarking techniques, maintaining ownership verification and traceability while preventing misuse of the model.

\bibliographystyle{ACM-Reference-Format}
\bibliography{sample-base}


\begin{thebibliography}{36}


\ifx \showCODEN    \undefined \def \showCODEN     #1{\unskip}     \fi
\ifx \showISBNx    \undefined \def \showISBNx     #1{\unskip}     \fi
\ifx \showISBNxiii \undefined \def \showISBNxiii  #1{\unskip}     \fi
\ifx \showISSN     \undefined \def \showISSN      #1{\unskip}     \fi
\ifx \showLCCN     \undefined \def \showLCCN      #1{\unskip}     \fi
\ifx \shownote     \undefined \def \shownote      #1{#1}          \fi
\ifx \showarticletitle \undefined \def \showarticletitle #1{#1}   \fi
\ifx \showURL      \undefined \def \showURL       {\relax}        \fi
\providecommand\bibfield[2]{#2}
\providecommand\bibinfo[2]{#2}
\providecommand\natexlab[1]{#1}
\providecommand\showeprint[2][]{arXiv:#2}

\bibitem[Brooks et~al\mbox{.}(2023)]%
        {brooks2023instructpix2pix}
\bibfield{author}{\bibinfo{person}{Tim Brooks}, \bibinfo{person}{Aleksander Holynski}, {and} \bibinfo{person}{Alexei~A Efros}.} \bibinfo{year}{2023}\natexlab{}.
\newblock \showarticletitle{Instructpix2pix: Learning to follow image editing instructions}. In \bibinfo{booktitle}{\emph{Proceedings of the IEEE/CVF Conference on Computer Vision and Pattern Recognition}}. \bibinfo{pages}{18392--18402}.
\newblock


\bibitem[Cao et~al\mbox{.}(2024)]%
        {cao2024controllable}
\bibfield{author}{\bibinfo{person}{Pu Cao}, \bibinfo{person}{Feng Zhou}, \bibinfo{person}{Qing Song}, {and} \bibinfo{person}{Lu Yang}.} \bibinfo{year}{2024}\natexlab{}.
\newblock \showarticletitle{Controllable generation with text-to-image diffusion models: A survey}.
\newblock \bibinfo{journal}{\emph{arXiv preprint arXiv:2403.04279}} (\bibinfo{year}{2024}).
\newblock


\bibitem[Chakraborty et~al\mbox{.}(2020)]%
        {chakraborty2020hardware}
\bibfield{author}{\bibinfo{person}{Abhishek Chakraborty}, \bibinfo{person}{Ankit Mondai}, {and} \bibinfo{person}{Ankur Srivastava}.} \bibinfo{year}{2020}\natexlab{}.
\newblock \showarticletitle{Hardware-assisted intellectual property protection of deep learning models}. In \bibinfo{booktitle}{\emph{2020 57th ACM/IEEE Design Automation Conference (DAC)}}. IEEE, \bibinfo{pages}{1--6}.
\newblock


\bibitem[Ci et~al\mbox{.}(2025)]%
        {ci2025ringid}
\bibfield{author}{\bibinfo{person}{Hai Ci}, \bibinfo{person}{Pei Yang}, \bibinfo{person}{Yiren Song}, {and} \bibinfo{person}{Mike~Zheng Shou}.} \bibinfo{year}{2025}\natexlab{}.
\newblock \showarticletitle{Ringid: Rethinking tree-ring watermarking for enhanced multi-key identification}. In \bibinfo{booktitle}{\emph{European Conference on Computer Vision}}. Springer, \bibinfo{pages}{338--354}.
\newblock


\bibitem[Cox et~al\mbox{.}(2007)]%
        {cox2007digital}
\bibfield{author}{\bibinfo{person}{Ingemar Cox}, \bibinfo{person}{Matthew Miller}, \bibinfo{person}{Jeffrey Bloom}, \bibinfo{person}{Jessica Fridrich}, {and} \bibinfo{person}{Ton Kalker}.} \bibinfo{year}{2007}\natexlab{}.
\newblock \bibinfo{booktitle}{\emph{Digital watermarking and steganography}}.
\newblock \bibinfo{publisher}{Morgan kaufmann}.
\newblock


\bibitem[Darvish~Rouhani et~al\mbox{.}(2019)]%
        {darvish2019deepsigns}
\bibfield{author}{\bibinfo{person}{Bita Darvish~Rouhani}, \bibinfo{person}{Huili Chen}, {and} \bibinfo{person}{Farinaz Koushanfar}.} \bibinfo{year}{2019}\natexlab{}.
\newblock \showarticletitle{Deepsigns: An end-to-end watermarking framework for ownership protection of deep neural networks}. In \bibinfo{booktitle}{\emph{Proceedings of the twenty-fourth international conference on architectural support for programming languages and operating systems}}. \bibinfo{pages}{485--497}.
\newblock


\bibitem[Feng et~al\mbox{.}(2024)]%
        {feng2024aqualora}
\bibfield{author}{\bibinfo{person}{Weitao Feng}, \bibinfo{person}{Wenbo Zhou}, \bibinfo{person}{Jiyan He}, \bibinfo{person}{Jie Zhang}, \bibinfo{person}{Tianyi Wei}, \bibinfo{person}{Guanlin Li}, \bibinfo{person}{Tianwei Zhang}, \bibinfo{person}{Weiming Zhang}, {and} \bibinfo{person}{Nenghai Yu}.} \bibinfo{year}{2024}\natexlab{}.
\newblock \showarticletitle{AquaLoRA: Toward White-box Protection for Customized Stablerombach2021highresolution Diffusion Models via Watermark LoRA}.
\newblock \bibinfo{journal}{\emph{arXiv preprint arXiv:2405.11135}} (\bibinfo{year}{2024}).
\newblock


\bibitem[Fernandez et~al\mbox{.}(2023)]%
        {fernandez2023stable}
\bibfield{author}{\bibinfo{person}{Pierre Fernandez}, \bibinfo{person}{Guillaume Couairon}, \bibinfo{person}{Herv{\'e} J{\'e}gou}, \bibinfo{person}{Matthijs Douze}, {and} \bibinfo{person}{Teddy Furon}.} \bibinfo{year}{2023}\natexlab{}.
\newblock \showarticletitle{The stable signature: Rooting watermarks in latent diffusion models}. In \bibinfo{booktitle}{\emph{Proceedings of the IEEE/CVF International Conference on Computer Vision}}. \bibinfo{pages}{22466--22477}.
\newblock


\bibitem[Gal et~al\mbox{.}(2022)]%
        {gal2022image}
\bibfield{author}{\bibinfo{person}{Rinon Gal}, \bibinfo{person}{Yuval Alaluf}, \bibinfo{person}{Yuval Atzmon}, \bibinfo{person}{Or Patashnik}, \bibinfo{person}{Amit~H Bermano}, \bibinfo{person}{Gal Chechik}, {and} \bibinfo{person}{Daniel Cohen-Or}.} \bibinfo{year}{2022}\natexlab{}.
\newblock \showarticletitle{An image is worth one word: Personalizing text-to-image generation using textual inversion}.
\newblock \bibinfo{journal}{\emph{arXiv preprint arXiv:2208.01618}} (\bibinfo{year}{2022}).
\newblock


\bibitem[Heusel et~al\mbox{.}(2017)]%
        {heusel2017gans}
\bibfield{author}{\bibinfo{person}{Martin Heusel}, \bibinfo{person}{Hubert Ramsauer}, \bibinfo{person}{Thomas Unterthiner}, \bibinfo{person}{Bernhard Nessler}, {and} \bibinfo{person}{Sepp Hochreiter}.} \bibinfo{year}{2017}\natexlab{}.
\newblock \showarticletitle{Gans trained by a two time-scale update rule converge to a local nash equilibrium}.
\newblock \bibinfo{journal}{\emph{Advances in neural information processing systems}}  \bibinfo{volume}{30} (\bibinfo{year}{2017}).
\newblock


\bibitem[Kumari et~al\mbox{.}(2023)]%
        {kumari2023multi}
\bibfield{author}{\bibinfo{person}{Nupur Kumari}, \bibinfo{person}{Bingliang Zhang}, \bibinfo{person}{Richard Zhang}, \bibinfo{person}{Eli Shechtman}, {and} \bibinfo{person}{Jun-Yan Zhu}.} \bibinfo{year}{2023}\natexlab{}.
\newblock \showarticletitle{Multi-concept customization of text-to-image diffusion}. In \bibinfo{booktitle}{\emph{Proceedings of the IEEE/CVF conference on computer vision and pattern recognition}}. \bibinfo{pages}{1931--1941}.
\newblock


\bibitem[Lei et~al\mbox{.}(2024)]%
        {lei2024diffusetrace}
\bibfield{author}{\bibinfo{person}{Liangqi Lei}, \bibinfo{person}{Keke Gai}, \bibinfo{person}{Jing Yu}, {and} \bibinfo{person}{Liehuang Zhu}.} \bibinfo{year}{2024}\natexlab{}.
\newblock \showarticletitle{Diffusetrace: A transparent and flexible watermarking scheme for latent diffusion model}.
\newblock \bibinfo{journal}{\emph{arXiv preprint arXiv:2405.02696}} (\bibinfo{year}{2024}).
\newblock


\bibitem[Li et~al\mbox{.}(2024)]%
        {li2024controlar}
\bibfield{author}{\bibinfo{person}{Zongming Li}, \bibinfo{person}{Tianheng Cheng}, \bibinfo{person}{Shoufa Chen}, \bibinfo{person}{Peize Sun}, \bibinfo{person}{Haocheng Shen}, \bibinfo{person}{Longjin Ran}, \bibinfo{person}{Xiaoxin Chen}, \bibinfo{person}{Wenyu Liu}, {and} \bibinfo{person}{Xinggang Wang}.} \bibinfo{year}{2024}\natexlab{}.
\newblock \showarticletitle{Controlar: Controllable image generation with autoregressive models}.
\newblock \bibinfo{journal}{\emph{arXiv preprint arXiv:2410.02705}} (\bibinfo{year}{2024}).
\newblock


\bibitem[Liu et~al\mbox{.}(2023)]%
        {liu2023t2iw}
\bibfield{author}{\bibinfo{person}{An-An Liu}, \bibinfo{person}{Guokai Zhang}, \bibinfo{person}{Yuting Su}, \bibinfo{person}{Ning Xu}, \bibinfo{person}{Yongdong Zhang}, {and} \bibinfo{person}{Lanjun Wang}.} \bibinfo{year}{2023}\natexlab{}.
\newblock \showarticletitle{T2IW: Joint Text to Image \& Watermark Generation}.
\newblock \bibinfo{journal}{\emph{arXiv preprint arXiv:2309.03815}} (\bibinfo{year}{2023}).
\newblock


\bibitem[Loshchilov(2017)]%
        {loshchilov2017decoupled}
\bibfield{author}{\bibinfo{person}{I Loshchilov}.} \bibinfo{year}{2017}\natexlab{}.
\newblock \showarticletitle{Decoupled weight decay regularization}.
\newblock \bibinfo{journal}{\emph{arXiv preprint arXiv:1711.05101}} (\bibinfo{year}{2017}).
\newblock


\bibitem[Meng et~al\mbox{.}(2024)]%
        {meng2024latent}
\bibfield{author}{\bibinfo{person}{Zheling Meng}, \bibinfo{person}{Bo Peng}, {and} \bibinfo{person}{Jing Dong}.} \bibinfo{year}{2024}\natexlab{}.
\newblock \showarticletitle{Latent Watermark: Inject and Detect Watermarks in Latent Diffusion Space}.
\newblock \bibinfo{journal}{\emph{arXiv preprint arXiv:2404.00230}} (\bibinfo{year}{2024}).
\newblock


\bibitem[Min et~al\mbox{.}(2025)]%
        {min2025watermark}
\bibfield{author}{\bibinfo{person}{Rui Min}, \bibinfo{person}{Sen Li}, \bibinfo{person}{Hongyang Chen}, {and} \bibinfo{person}{Minhao Cheng}.} \bibinfo{year}{2025}\natexlab{}.
\newblock \showarticletitle{A watermark-conditioned diffusion model for ip protection}. In \bibinfo{booktitle}{\emph{European Conference on Computer Vision}}. Springer, \bibinfo{pages}{104--120}.
\newblock


\bibitem[Namba and Sakuma(2019)]%
        {namba2019robust}
\bibfield{author}{\bibinfo{person}{Ryota Namba} {and} \bibinfo{person}{Jun Sakuma}.} \bibinfo{year}{2019}\natexlab{}.
\newblock \showarticletitle{Robust watermarking of neural network with exponential weighting}. In \bibinfo{booktitle}{\emph{Proceedings of the 2019 ACM Asia Conference on Computer and Communications Security}}. \bibinfo{pages}{228--240}.
\newblock


\bibitem[Peng et~al\mbox{.}(2025)]%
        {peng2025intellectual}
\bibfield{author}{\bibinfo{person}{Sen Peng}, \bibinfo{person}{Yufei Chen}, \bibinfo{person}{Cong Wang}, {and} \bibinfo{person}{Xiaohua Jia}.} \bibinfo{year}{2025}\natexlab{}.
\newblock \showarticletitle{Intellectual property protection of diffusion models via the watermark diffusion process}. In \bibinfo{booktitle}{\emph{International Conference on Web Information Systems Engineering}}. Springer, \bibinfo{pages}{290--305}.
\newblock


\bibitem[Ramesh et~al\mbox{.}(2022)]%
        {ramesh2022hierarchical}
\bibfield{author}{\bibinfo{person}{Aditya Ramesh}, \bibinfo{person}{Prafulla Dhariwal}, \bibinfo{person}{Alex Nichol}, \bibinfo{person}{Casey Chu}, {and} \bibinfo{person}{Mark Chen}.} \bibinfo{year}{2022}\natexlab{}.
\newblock \showarticletitle{Hierarchical text-conditional image generation with clip latents}.
\newblock \bibinfo{journal}{\emph{arXiv preprint arXiv:2204.06125}} \bibinfo{volume}{1}, \bibinfo{number}{2} (\bibinfo{year}{2022}), \bibinfo{pages}{3}.
\newblock


\bibitem[Rombach et~al\mbox{.}(2021)]%
        {rombach2021highresolution}
\bibfield{author}{\bibinfo{person}{Robin Rombach}, \bibinfo{person}{Andreas Blattmann}, \bibinfo{person}{Dominik Lorenz}, \bibinfo{person}{Patrick Esser}, {and} \bibinfo{person}{Björn Ommer}.} \bibinfo{year}{2021}\natexlab{}.
\newblock \bibinfo{title}{High-Resolution Image Synthesis with Latent Diffusion Models}.
\newblock
\showeprint[arxiv]{2112.10752}~[cs.CV]


\bibitem[Rombach et~al\mbox{.}(2022a)]%
        {rombach2022high}
\bibfield{author}{\bibinfo{person}{Robin Rombach}, \bibinfo{person}{Andreas Blattmann}, \bibinfo{person}{Dominik Lorenz}, \bibinfo{person}{Patrick Esser}, {and} \bibinfo{person}{Bj{\"o}rn Ommer}.} \bibinfo{year}{2022}\natexlab{a}.
\newblock \showarticletitle{High-resolution image synthesis with latent diffusion models}. In \bibinfo{booktitle}{\emph{Proceedings of the IEEE/CVF conference on computer vision and pattern recognition}}. \bibinfo{pages}{10684--10695}.
\newblock


\bibitem[Rombach et~al\mbox{.}(2022b)]%
        {Rombach_2022_CVPR}
\bibfield{author}{\bibinfo{person}{Robin Rombach}, \bibinfo{person}{Andreas Blattmann}, \bibinfo{person}{Dominik Lorenz}, \bibinfo{person}{Patrick Esser}, {and} \bibinfo{person}{Bj\"orn Ommer}.} \bibinfo{year}{2022}\natexlab{b}.
\newblock \showarticletitle{High-Resolution Image Synthesis With Latent Diffusion Models}. In \bibinfo{booktitle}{\emph{Proceedings of the IEEE/CVF Conference on Computer Vision and Pattern Recognition (CVPR)}}. \bibinfo{pages}{10684--10695}.
\newblock


\bibitem[Ruiz et~al\mbox{.}(2023)]%
        {ruiz2023dreambooth}
\bibfield{author}{\bibinfo{person}{Nataniel Ruiz}, \bibinfo{person}{Yuanzhen Li}, \bibinfo{person}{Varun Jampani}, \bibinfo{person}{Yael Pritch}, \bibinfo{person}{Michael Rubinstein}, {and} \bibinfo{person}{Kfir Aberman}.} \bibinfo{year}{2023}\natexlab{}.
\newblock \showarticletitle{Dreambooth: Fine tuning text-to-image diffusion models for subject-driven generation}. In \bibinfo{booktitle}{\emph{Proceedings of the IEEE/CVF conference on computer vision and pattern recognition}}. \bibinfo{pages}{22500--22510}.
\newblock


\bibitem[Ruiz et~al\mbox{.}(2024)]%
        {ruiz2024hyperdreambooth}
\bibfield{author}{\bibinfo{person}{Nataniel Ruiz}, \bibinfo{person}{Yuanzhen Li}, \bibinfo{person}{Varun Jampani}, \bibinfo{person}{Wei Wei}, \bibinfo{person}{Tingbo Hou}, \bibinfo{person}{Yael Pritch}, \bibinfo{person}{Neal Wadhwa}, \bibinfo{person}{Michael Rubinstein}, {and} \bibinfo{person}{Kfir Aberman}.} \bibinfo{year}{2024}\natexlab{}.
\newblock \showarticletitle{Hyperdreambooth: Hypernetworks for fast personalization of text-to-image models}. In \bibinfo{booktitle}{\emph{Proceedings of the IEEE/CVF conference on computer vision and pattern recognition}}. \bibinfo{pages}{6527--6536}.
\newblock


\bibitem[Shoshan et~al\mbox{.}(2021)]%
        {shoshan2021gan}
\bibfield{author}{\bibinfo{person}{Alon Shoshan}, \bibinfo{person}{Nadav Bhonker}, \bibinfo{person}{Igor Kviatkovsky}, {and} \bibinfo{person}{Gerard Medioni}.} \bibinfo{year}{2021}\natexlab{}.
\newblock \showarticletitle{Gan-control: Explicitly controllable gans}. In \bibinfo{booktitle}{\emph{Proceedings of the IEEE/CVF international conference on computer vision}}. \bibinfo{pages}{14083--14093}.
\newblock


\bibitem[Sohn et~al\mbox{.}(2023)]%
        {sohn2023styledrop}
\bibfield{author}{\bibinfo{person}{Kihyuk Sohn}, \bibinfo{person}{Nataniel Ruiz}, \bibinfo{person}{Kimin Lee}, \bibinfo{person}{Daniel~Castro Chin}, \bibinfo{person}{Irina Blok}, \bibinfo{person}{Huiwen Chang}, \bibinfo{person}{Jarred Barber}, \bibinfo{person}{Lu Jiang}, \bibinfo{person}{Glenn Entis}, \bibinfo{person}{Yuanzhen Li}, {et~al\mbox{.}}} \bibinfo{year}{2023}\natexlab{}.
\newblock \showarticletitle{Styledrop: Text-to-image generation in any style}.
\newblock \bibinfo{journal}{\emph{arXiv preprint arXiv:2306.00983}} (\bibinfo{year}{2023}).
\newblock


\bibitem[Tancik et~al\mbox{.}(2020)]%
        {tancik2020stegastamp}
\bibfield{author}{\bibinfo{person}{Matthew Tancik}, \bibinfo{person}{Ben Mildenhall}, {and} \bibinfo{person}{Ren Ng}.} \bibinfo{year}{2020}\natexlab{}.
\newblock \showarticletitle{Stegastamp: Invisible hyperlinks in physical photographs}. In \bibinfo{booktitle}{\emph{Proceedings of the IEEE/CVF conference on computer vision and pattern recognition}}. \bibinfo{pages}{2117--2126}.
\newblock


\bibitem[Wang et~al\mbox{.}(2021)]%
        {wang2021realesrgan}
\bibfield{author}{\bibinfo{person}{Xintao Wang}, \bibinfo{person}{Liangbin Xie}, \bibinfo{person}{Chao Dong}, {and} \bibinfo{person}{Ying Shan}.} \bibinfo{year}{2021}\natexlab{}.
\newblock \showarticletitle{Real-ESRGAN: Training Real-World Blind Super-Resolution with Pure Synthetic Data}. In \bibinfo{booktitle}{\emph{International Conference on Computer Vision Workshops (ICCVW)}} (2021).
\newblock


\bibitem[Wang et~al\mbox{.}(2004)]%
        {wang2004image}
\bibfield{author}{\bibinfo{person}{Zhou Wang}, \bibinfo{person}{Alan~C Bovik}, \bibinfo{person}{Hamid~R Sheikh}, {and} \bibinfo{person}{Eero~P Simoncelli}.} \bibinfo{year}{2004}\natexlab{}.
\newblock \showarticletitle{Image quality assessment: from error visibility to structural similarity}.
\newblock \bibinfo{journal}{\emph{IEEE transactions on image processing}} \bibinfo{volume}{13}, \bibinfo{number}{4} (\bibinfo{year}{2004}), \bibinfo{pages}{600--612}.
\newblock


\bibitem[Wen et~al\mbox{.}(2023)]%
        {wen2023tree}
\bibfield{author}{\bibinfo{person}{Yuxin Wen}, \bibinfo{person}{John Kirchenbauer}, \bibinfo{person}{Jonas Geiping}, {and} \bibinfo{person}{Tom Goldstein}.} \bibinfo{year}{2023}\natexlab{}.
\newblock \showarticletitle{Tree-ring watermarks: Fingerprints for diffusion images that are invisible and robust}.
\newblock \bibinfo{journal}{\emph{arXiv preprint arXiv:2305.20030}} (\bibinfo{year}{2023}).
\newblock


\bibitem[Xiong et~al\mbox{.}(2023)]%
        {xiong2023flexible}
\bibfield{author}{\bibinfo{person}{Cheng Xiong}, \bibinfo{person}{Chuan Qin}, \bibinfo{person}{Guorui Feng}, {and} \bibinfo{person}{Xinpeng Zhang}.} \bibinfo{year}{2023}\natexlab{}.
\newblock \showarticletitle{Flexible and secure watermarking for latent diffusion model}. In \bibinfo{booktitle}{\emph{Proceedings of the 31st ACM International Conference on Multimedia}}. \bibinfo{pages}{1668--1676}.
\newblock


\bibitem[Yang et~al\mbox{.}(2024)]%
        {yang2024gaussian}
\bibfield{author}{\bibinfo{person}{Zijin Yang}, \bibinfo{person}{Kai Zeng}, \bibinfo{person}{Kejiang Chen}, \bibinfo{person}{Han Fang}, \bibinfo{person}{Weiming Zhang}, {and} \bibinfo{person}{Nenghai Yu}.} \bibinfo{year}{2024}\natexlab{}.
\newblock \showarticletitle{Gaussian Shading: Provable Performance-Lossless Image Watermarking for Diffusion Models}. In \bibinfo{booktitle}{\emph{Proceedings of the IEEE/CVF Conference on Computer Vision and Pattern Recognition}}. \bibinfo{pages}{12162--12171}.
\newblock


\bibitem[Zhang et~al\mbox{.}(2019)]%
        {zhang2019robust}
\bibfield{author}{\bibinfo{person}{Kevin~Alex Zhang}, \bibinfo{person}{Lei Xu}, \bibinfo{person}{Alfredo Cuesta-Infante}, {and} \bibinfo{person}{Kalyan Veeramachaneni}.} \bibinfo{year}{2019}\natexlab{}.
\newblock \showarticletitle{Robust invisible video watermarking with attention}.
\newblock \bibinfo{journal}{\emph{arXiv preprint arXiv:1909.01285}} (\bibinfo{year}{2019}).
\newblock


\bibitem[Zhang et~al\mbox{.}(2018)]%
        {zhang2018unreasonable}
\bibfield{author}{\bibinfo{person}{Richard Zhang}, \bibinfo{person}{Phillip Isola}, \bibinfo{person}{Alexei~A Efros}, \bibinfo{person}{Eli Shechtman}, {and} \bibinfo{person}{Oliver Wang}.} \bibinfo{year}{2018}\natexlab{}.
\newblock \showarticletitle{The unreasonable effectiveness of deep features as a perceptual metric}. In \bibinfo{booktitle}{\emph{Proceedings of the IEEE conference on computer vision and pattern recognition}}. \bibinfo{pages}{586--595}.
\newblock


\bibitem[Zhu et~al\mbox{.}(2018)]%
        {zhu2018hidden}
\bibfield{author}{\bibinfo{person}{Jiren Zhu}, \bibinfo{person}{Russell Kaplan}, \bibinfo{person}{Justin Johnson}, {and} \bibinfo{person}{Li Fei-Fei}.} \bibinfo{year}{2018}\natexlab{}.
\newblock \showarticletitle{Hidden: Hiding data with deep networks}. In \bibinfo{booktitle}{\emph{Proceedings of the European conference on computer vision (ECCV)}}. \bibinfo{pages}{657--672}.
\newblock


\end{thebibliography}

\end{document}